# Interstitial flow potentiates TGF-$\beta$/Smad-signaling activity in lung cancer spheroids in a 3D-microfluidic chip†

Zaid Rahman,[a] Ankur Bordoloi,[a] Haifa Rouhana,[a] Margherita Tavasso,[a] Gerard van der Zon,[b] Valeria Garbin,[a] Peter ten Dijke,[b] and Pouyan E. Boukany *[a]

Within the tumor microenvironment (TME), cancer cells use mechanotransduction pathways to convert biophysical forces to biochemical signals. However, the underlying mechanisms and functional significance of these pathways remain largely unclear. The upregulation of mechanosensitive pathways from biophysical forces such as interstitial flow, leads to the activation of various cytokines, including transforming growth factor-β (TGF-β). TGF-β is a critical inducer of epithelial-mesenchymal transition (EMT) in cancer cells that leads to increased cell motility and invasion in a Smad dependent manner. Current research models have limited ability to investigate the combined effects of biophysical forces (such as interstitial flow) and cytokines (TGF-β) in a 3D microenvironment. We used a 3D-matrix based microfluidic platform to demonstrate the potentiating effect of interstitial flow (IF) on exogenous TGF-β induced upregulation of the Smad-signaling activity and the expression of mesenchymal marker vimentin in A549 lung cancer spheroids. To monitor this, we used stably integrated fluorescent based reporters into the A549 cancer cell genome. Our results demonstrate that interstitial flow enhances exogenous TGF-β induced Smad-signaling activity in lung cancer spheroids embedded in a matrix microenvironment. In addition, we observed an increased cell motility for A549 spheroids when exposed to interstitial flow and TGF-β. Our 3D-microfluidic model integrated with real-time imaging provides a powerful tool for investigating cancer cell signaling and motility associated with invasion characteristics in a physiologically relevant TME.

## 1 Introduction

The 3D tumor microenvironment (TME) plays a crucial role in the progression and metastasis of primary tumors to secondary tumor sites [1,2]. It consists of key components such as the extracellular matrix (ECM), biophysical forces (interstitial flow and consequent fluid stresses), tumor cell-TME interactions in the presence of stromal cells, immune cells and cancer-associated fibroblasts (CAFs). The interplay of these components contributes to the metastatic cascade of events from early dissemination to extravasation [3]. However, most tumor cell migration and invasion studies have been performed in 2D/3D in-vitro models that poorly recapitulate the characteristics of solid tumors in-vivo. To overcome these limitations, microfluidic platforms provide an effective tool to replicate a physiologically relevant TME for studying cancer cell behavior [4].

Recent advances in microfluidic platforms based on a 3-D matrix have allowed for the incorporation of key components of the tumor microenvironment (TME) in cancer cell migration and invasion studies [4,5]. To mimic the ECM, natural-hydrogel materials with tunable mechanical properties are used [6]. These hydrogels have been further embedded with single cancer cells and/or cancer cell aggregate/spheroids to include cell-matrix interactions. Advancements in modeling and fabrication technologies have improved microfluidic devices to introduce interstitial flow (IF) for long-term perfusion and culture conditions. In the past, IF studies were mostly performed on single cells embedded in a matrix material [7–9]. Recently, researchers have investigated the invasive and migratory cellular response of a breast tumor spheroid model under IF to show morphological and epigenetic changes [10]. However, these studies were limited to highly migratory breast cancer cells and did not include effect of biochemical signals towards EMT signaling pathways.

The role of IF is important due to its direct influence on the remodeling of the ECM, where compressive, tensional and shear forces are sensed by cell-surface receptors that activate mechanotransduction pathways to trigger biochemical signals [7,11–13].

[a] *Department of Chemical Engineering, Delft University of Technology, Delft, The Netherlands. E-mail: p.e.boukany@tudelft.nl*
[b] *Department of Cell and Chemical Biology and Oncode Institute, Leiden University Medical Center, Leiden, The Netherlands.*

† Electronic Supplementary Information (ESI) available: See DOI: 00.0000/00000000.

This further leads to the activation and upregulation of many core EMT cytokines [7], including the TGF-β cytokine, known as a key EMT inducer [14,15]. In solid tumors, the poorly drained interstitial flow is responsible for interstitial fluid pressure build up in the surrounding healthy tissue [7]. Moreover, the lung tumor tissue is constantly subjected to a mechanical load due to its physiological activities that may aid in cancer cell invasion and migration [16,17]. Therefore, it is of primary interest to study primary lung tumor models such as A549 lung adenocarcinoma, when subjected to biophysical force induced stresses. It has been proposed that the cancer cells exposed to biomechanical forces (such as IF, fluid-induced shear stress, compressive stress from matrix microenvironment) leads to endogenous TGF-β driven Smad-signaling activity towards EMT response [13,18,19]. Moreover, studies have also investigated the role of fluid-induced shear stress to promote mechanotransduction pathways (such as YAP/TAZ) responsible for triggering EMT signaling for cancer cell invasion in non-small cell lung cancer, breast cancer and melanoma tumor [20–22].

TGF-β is capable to promote cancer cell invasion and progression in various tumor types such as lung, breast and pancreatic cancer [23,24]. TGF-β receptors at the cell-surface upon binding TGF-β activate the intracellular Smad-signaling pathway [23]. Activated Smads can act as transcription factors to mediate EMT associated with cancer. Many researchers have studied the role of TGF-β in static 2D/3D tumor models, highlighting its importance in activating EMT transcriptional factors including SNAI1, TWIST, ZEB1[23]. Studies conducted on A549 lung adenocarcinoma cells showed EMT behavior upon exposure to TGF-β cytokine [25–28]. Most studies focused on the upregulation of mesenchymal markers (such as vimentin) and an increased expression of transcription factors such as SNAIl and ZEB2 highlighting EMT response [25,29,30]. The upregulation of the vimentin mesenchymal marker and downregulation of the E-cadherin (epithelial marker) in A549 lung cancer cells were found to be associated with an aggressive motile response [24]. In recent years, researchers further studied A549 3D cancer models towards EMT behavior [31,32]. However, these studies were performed in culture conditions devoid of matrix material and IF. Thus, there is an evident lack of research on the effect of IF and exogenous TGF-β on A549 lung tumor EMT response in a relevant matrix microenvironment.

In this research, we employed a 3D-matrix based microfluidic model to investigate the impact of IF and exogenous TGF-β cytokine on epithelial-like A549 spheroids (Fig. 1(A) and (B)). Specifically, we investigated the Smad-dependent transcriptional pathway and vimentin biomarker expression in response to varying IF and exogenous TGF-β concentration towards cancer cell invasion(Fig. 1(C)). These studies were conducted with genetically modified A549 lung tumor cells with dual artificial reporter constructs for Smad-signaling pathway (CAGA-12-GFP reporter gene) and vimentin biomarker (VIM-RFP reporter gene) (Fig. 1C, inset figure). We demonstrate that IF potentiates Smad-dependent transcriptional response when exposed to exogenous TGF-β. The combined effect of IF and TGF-β also showed increased abundance in vimentin protein. Lastly, A549 lung tumor exhibited increased cellular motion on the spheroid periphery showing cancer cell invasion characteristics. These findings suggest that external IF and cellular cues play critical roles in promoting the invasive characteristics of cancer cells within relevant matrix microenvironments, and highlight the importance of incorporating these factors in cancer research models.

## 2 Materials and Methods

### 2.1 Cell culture

A549-VIM-RFP cells were acquired from the company, ATCC [33]. These cells are engineered by CRISPR to produce a red fluorescent protein (RFP)-vimentin fusion protein. When cells acquire mesenchymal phenotype they express RFP linked to vimentin protein (VIM-RFP). To construct a dual reporter, A549-VIM-RFP cells were transduced with a lentiviral CAGA-12-GFP construct to produce a green fluorescent protein (GFP) response upon Smad-pathway activation. The dual reporter cell line was a gift from Yifan Zhu (Department of Cell and Chemical Biology, LUMC). The functionality of the CAGA-12-GFP reporter has been validated through several studies. [34,35]. A549-VIM-RFP cells have been previously reported to exhibit EMT with an upregulation in VIM-RFP fluorescence upon exogenous TGF-β stimulation [36]. The dual-reporter A549 cells were maintained in Dulbecco's Modified Eagle Medium High Glucose (DMEM,Sigma) containing 4.5 g/L glucose, L-glutamine without sodium pyruvate, and supplemented with 10% Fetal Bovine Serum (FBS, Sigma) and 1% Antibiotic-Antimycotic solution (Gibco). All cells were incubated at 37ºC with 5% $CO_2$ and sub-cultured 2 times per week. Cells were frequently tested for absence of mycoplasma and checked for authenticity by STR profiling.

### 2.2 Spheroid fabrication

Spheroids were grown in a commercially available Corning$^{TM}$ Elplasia$^{TM}$ 96-well plate for high-throughput spheroid production. These well plates are round-bottom with Ultra-Low Attachment (ULA) surface that prevents cell-surface attachment and promotes cell-cell adhesion. We used an initial seeding density of $40 \times 10^3$ cells (500 cells per micro-well) for each well to produce 79 spheroids. Spheroid size is dependent on the initial seeding density, cell proliferation rate and culture duration. Spheroids were ready to use after 4 days of culture in the wells and were $200 \pm 20\,\mu$m in diameter. We restricted the spheroid diameter to less than 220 $\mu$m to avoid a necrotic core and to avoid contact with the glass bottom of the microfluidic chip. Any cancer spheroids that made contact with the microfluidic glass substrate, were excluded from the analysis in this work.

### 2.3 Hydrogel synthesis and characterization

Gelatin methacryloyl (GelMA), 300g bloom, 60 % degree substitution, was purchased from Sigma Aldrich. Like gelatin, gelMA is still a thermo-reversible gel, however, the methacrylic anhydride groups give the ability to undergo covalent cross-linking under UV light (365 nm) in the presence of a UV photo-initiator. 5wt.% gelMA was used in experiments, with a mass ratio of 1:16 of photo-initiator (Lithium phenyl-2,4,6- trimethylbenzoylphosphinate, LAP; Sigma Aldrich). LAP and gelMA were added together and dissolved in Dulbecco's Phosphae Buffered Saline (DPBS; Gibco). The mixture was dissolved at 37°C in a water bath for about 2 hours. The hydrogel was then crosslinked using Colibri Axio Observer microscope laser 385 nm with a 5x objective lens for 45 seconds. The viscoelastic properties of crosslinked GelMA were investigated with a modular rotational rheometer (DSR 502, Anton Paar) equipped with a parallel plate of a diameter of 25 mm. Full experimental detail can be found in ESI (section 1). The 5wt.% GelMA analyzed at a fixed strain of 1% with frequency sweeps (0.1 to 100 rad/s) at room temperature showed a solid-like behavior, with a storage modulus G'$\approx$ 250 Pa, higher than the loss modulus (G") by at least one order of magnitude (Fig. S1). The lung tumor tissue stiffness is reported around 200 Pa in vivo [37]. To replicate the mechanical properties of TME (under in vivo conditions), we employed a matrix material with similar mechanical properties.

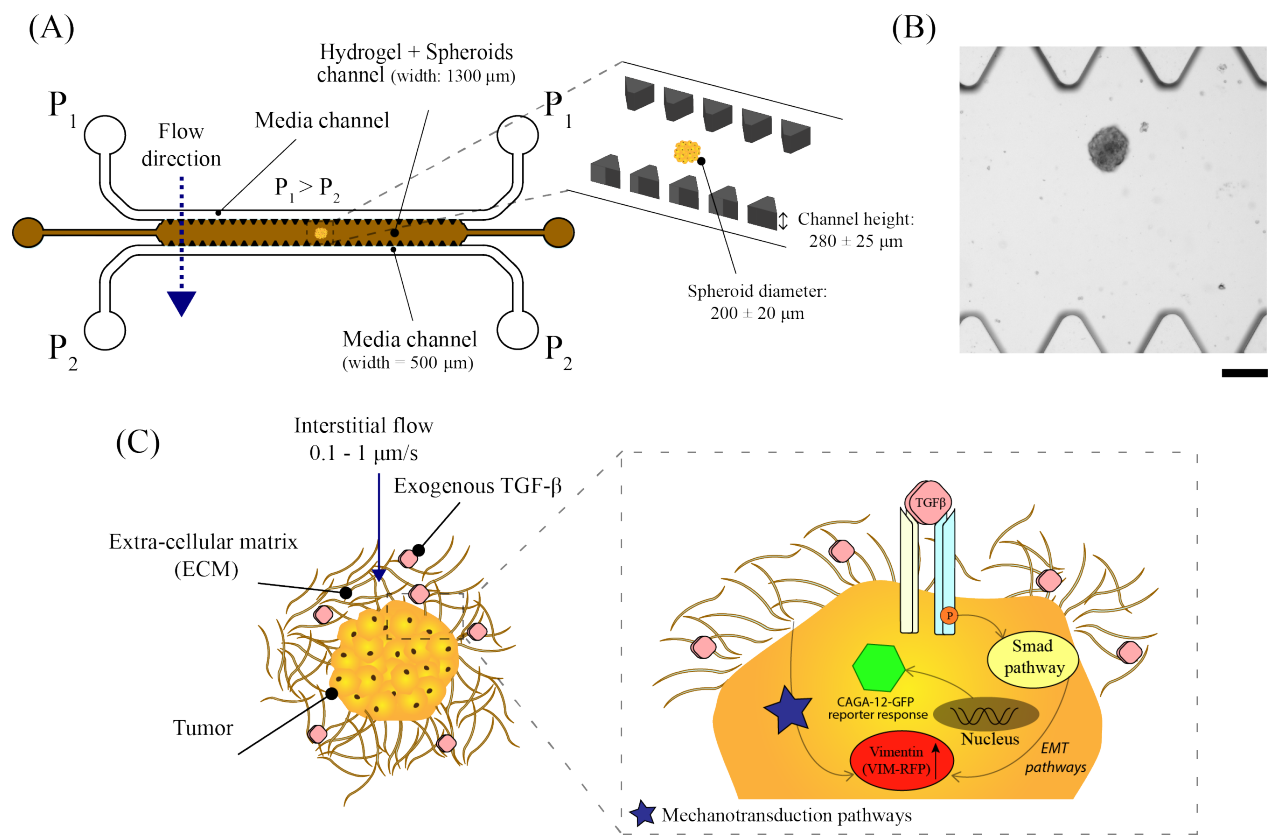

Fig. 1 3D-matrix based microfluidic platform to study interstitial flow and TGF-$\beta$/Smad-signaling and vimentin expression in A549 lung tumor spheroids. (A) Schematic of the 3D-matrix based microfluidic platform. Inset figure shows the spheroid size and the dimensions of the channel height of the microfluidic chip (not to scale). (B) Bright-field image displaying A549 spheroid embedded in 3D-matrix based scaffold in the hydrogel channel of the microfluidic chip. Scale bar: 200 $\mu$m (C) A549 lung tumor spheroid exposed to interstitial flow and exogenous TGF-$\beta$ embedded in a matrix microenvironment. Inset figure illustrates that exogenous TGF-$\beta$ molecules specifically binds with TGF-$\beta$ surface receptors to activate the intracellular Smad-signalling pathway. This results in upregulation of transcriptional reporter gene CAGA-12-GFP activity. TGF-$\beta$ cytokine also leads to the upregulation in EMT pathways. Upregulation in EMT biomarker (vimentin) can be investigated by determining the VIM-RFP reporter intensity. Moreover, cancer cells sense interstitial flow induced fluid-shear and hydrodynamic stress to initiate mechanotransduction pathways triggering EMT. These biomechanical forces may enforce the TGF-$\beta$/Smad activity for increased transcriptional reporter activity (CAGA-12-GFP intensity) and EMT response (VIM-RFP reporter expression).

### 2.4 Microfluidic chip fabrication and interstitial flow characterization

The microfluidic chip was fabricated on a 4-inch silicon wafer by the photo-lithography process in a cleanroom facility using $\mu$MLA Laser Writer (Heidelberg Instruments) (full procedure described in ESI, section 2). The microfluidic chip design was inspired by IF studies performed with single cells and was upgraded to fabricate a channel height of 280 $\pm$ 25 $\mu$m (Fig.1A) [38–40]. From the master mould, polydimethylsiloxane (PDMS) based microfluidic chips were fabricated by soft-lithography technique (refer to ESI, section 2 for a detailed procedure).

The microfluidic chip consisted of three parallel channels separated by triangular pillars (all side lengths: 150 $\mu$m and height: 280 $\pm$ 25 $\mu$m). The middle channel is loaded with 5 wt.% gelMA hydrogel, which is crosslinked under UV-light (385 nm) for 45 seconds. The top and bottom channels are the fluidic channels. The inlets of the top channel were maintained at a higher pressure ($P_1$) relative to the bottom channel ($P_2$) to generate an IF along the pressure gradient (Fig.1A). By controlling the pressure of the reservoirs at ($P_1$) and ($P_2$), we were able to establish a pres-sure gradient to generate an IF through gelMA hydrogel across the microfluidic device. The inlet and outlet pressures were controlled by a pressure pump (Fluigent) and operated via InFlow software to pressurize the sample reservoirs. Fig.S2 shows the experimental setup for generating a continuous IF using a pressure pump device connected to the 3D-microfluidic chip. According to Darcy's Law, flow velocity through a porous material is directly proportional to the pressure gradient governed by hydraulic permeability (K) of the material. In this case, we first calculated the hydraulic permeability of 5wt.% gelMA and then estimated the average IF velocity ($u_m$); refer to ESI, section 3 for detailed protocol. We tested two pressure drops ($\Delta P = P_1 - P_2$) of 20 mbar and 30 mbar that corresponded to an interstitial flow velocity of $u_m = 0.2$ μm/s and 0.45 μm/s obtained via COMSOL Multiphysics using Free and Porous Media Flow interface (Fig.S3, A and B). The IF velocity calculated for our 3D-matrix based microfluidic system is physiologically relevant as previously reported. Interstitial flow velocity in tumor tissues, performed in-vivo, in-vitro or via mathematical modelling are reported in the order of 0.01 - 1 μm/s in various cancer types [7,41,42]. Moreover, several studies have highlighted the role of elevated interstitial fluid pressure (IFP) in a tumor tissue as a barrier to tumor treatment [43]. IFP is reported in the order of 10 mbar to 60 mbar in various cancer types such as breast and melanoma skin cancer [43] and other studies (modeling and in-vivo experiments) has reported from 1 - 100 mbar [41,44].

### 2.5 Microfluidic device setup for IF and exogenous TGF-β studies

To investigate the effect of IF and exogenous TGF-β on spheroids, we used a step-wise procedure as described below. We first collected A549 spheroids from an Elplasia 96-well plate after 4 days of culture duration. The collected spheroids were then transferred to an empty well of a separate Corning Ultra-Low Attachment (ULA) 96-well plate. Once all the spheroids settled at the bottom of the well after 5 minutes, the cell culture media was aspirated out leaving only the spheroids in the well. A small volume of 5wt.% gelMA was added to this well to make a hydrogel-spheroid suspension. The hydrogel-spheroid suspension was then pipetted into the middle channel of the microfluidic device allowing entry of multiple spheroids. Once the middle channel was full, we gently removed the pipette from the inlet without introducing any air bubbles. The chip was then transferred to a microscope stage for UV-crosslinking at 385 nm laser source for 45 seconds using a 5x objective lens. After UV-irradiation, the hydrogel undergoes irreversible chemical crosslinking and acts as a 3D scaffold for spheroids (Fig.1B). To generate IF, we operated the microfluidic device as described in section 2.4. For experiments to study the effect of interstitial flow on A549 spheroids, the sample reservoir for the top channel was replaced with cell culture medium (DMEM, high glucose, 10% v/v FBS, 1% v/v Antibiotics). For IF with exogenous TGF-β experiments on A549 spheroids, we supplemented the culture medium with exogenous TGF-β (stock concentration; 5μg/mL) to achieve a final concentration of 0.1-10 ng/mL. Brightfield and fluorescent images of the spheroids were captured on an inverted fluorescence microscope (Zeiss Axio-Observer) at an interval of 1 hour for a duration of 70 hours using a 20x/NA 0.16 air objective and ORCA Flash 4.0 V2 (Hamamatsu) digital camera with a resolution of 2048x2048 pixels. We used Software Autofocus strategy with best contrast method to reduce background or out of focus fluorescence signal. For the GFP and RFP fluorescence, we used the 488 LED source (ex: 488 nm; emm: 520 nm) and 543 LED source (ex: 543 nm; emm: 590 nm), respectively. All experiments were conducted at 37ºC and 5% $CO_2$ using a stage top incubator (ibidi). Bright field images were taken at 10% light intensity and 100 millisecond exposure time. Fluorescent signal intensity for GFP and RFP images were analyzed via ImageJ (v1.53t, National Institute of Health, USA). A region of interest was created encircling the entire spheroid area for both GFP and RFP channel images, performed separately. This region of interest was quantified for pixel intensity density at every time point using Measure function in ImageJ. The fluorescent intensity signal values were normalized with respect to the signal intensity at t = 0 hr. CAGA-12-GFP and RFP reporter expression was plotted for multiple spheroids performed in 2 or 3 independent experiments. The device is robustly operational at pressure differences upto 30-35 mbar in the presence of spheroids. Increasing the pressure drop, resulted in the hydrogel structure breaking and interrupted uniform IF after a few hours. Within this pressure drop range, we were able to perform long-term culture experiments (up to 70 hours) to visualize cancer cell spheroid for fluorescence reporter signaling activity, and invasive response.

### 2.6 Microfluidic device setup for 2D cultured A549 cells under flow

Since A549 spheroids embedded in gelMA in a 3D-microfluidic chip cannot be retrieved to perform qPCR for target gene analyses, additional experiments were performed using dual-reporter A549 cells cultured in 2D-microfluidic without hydrogel matrix (see, Fig.S4). These 2D-microfluidic experiments (without matrix) enabled us to extract A549 cells after the experiment to run qPCR analyses on TGF-β and EMT target genes to complement CAGA-12-GFP and VIM-RFP reporter expression quantification. Full experimental procedure is described in ESI section 5. These studies additionally provide evidence of CAGA-12-GFP and VIM-RFP reporter expression when A549 single cells were exposed to 2D-flow (without matrix) alone and in combination with exogenous TGF-β conditions, shown in Fig.S5(A and B). Reporter expression was quantified in the following conditions: A) No-flow No-TGF-β (control), B) No-flow + TGF-β , C) Flow - No-TGF-β , and D) Flow + TGF-β conditions, shown in Fig.S5 (C and D) for CAGA-12-GFP and VIM-RFP reporter respectively.

### 2.7 qPCR analyses on target gene expression

Experiments performed using A549 cells in 2D-microfluidics (without matrix) in ESI section 5, were further used to establish CAGA-12-GFP reporter activity with TGF-β target genes and confirm VIM-RFP expression with EMT target genes (full experimental procedure described in ESI section 6.) We performed qPCR analyses on *CTGF, Serpin* (encoding PAI-1) and *Smad7* for TGF-β target genes and *E-cadherin, N-cadherin* and *Vimentin* for EMT target genes. Fig.S6 and S7 shows the relative change in mRNA expression for each condition with respect to No-flow No TGF-β (control) for TGF-β and EMT target genes respectively. qPCR analyses further helped to establish the effect of 2D-flow and/or exogenous TGF-β induced reporter expression at a molecular level.

## 2.8 Statistical analysis

All statistical analysis was performed using Microsoft Excel (Microsoft Corporation, USA). The statistical significant differences between the two experimental groups were determined by Student t-test using the function *t-test: two samples with unequal variance* and p values below 0.05 were considered to be significant. We categorize statistical differences as following; $p < 0.001$ (∗∗∗), $p < 0.01$ (∗∗) and $p < 0.05$ (∗).

# 3 Results and discussion

## 3.1 Exogenous TGF-β induced CAGA-12-GFP reporter response under interstitial flow conditions

To analyze the effect of exogenous TGF-β under interstitial flow (IF) on Smad3/4-dependent transcriptional reporter response, we first examined the overall CAGA-12-GFP reporter fluorescence intensities at the end of 70 hrs for a fixed $C_0 = 10$ ng/ml of exogenous TGF-β : (i) with IF (IF$^+$TGF-β $^+$) and (ii) without IF (IF$^-$TGF-β $^+$). These two conditions are contrasted with an IF condition without any exogenous TGF-β (IF$^+$TGF-β $^-$). Fig. 2A shows the brightfield images superposed with GFP fluorescence intensity at t = 0 and 70 hrs for these three conditions. The IF conditions were obtained under a fixed pressure gradient of $\Delta P = 30$ mbar, equivalent of an average interstitial fluid velocity, $u_m = 0.45$ μm/s (measured separately via an independent experiment; see Fig. S3.B). We observed an enhanced CAGA-12-GFP reporter expression with the addition of exogenous TGF-β (2A (ii) vs. (iii)), which becomes further amplified across the spheroid under the imposed IF (2A (i) vs. (ii)). This observation strongly suggests that IF enhances the exogenous TGF-β induced Smad-signaling activity in A549 spheroids.

The statistics of relative increase in the CAGA-12-GFP reporter expression ($I_{70}$) at t=70 hrs for these conditions were quantified for multiple spheroids based on the intensity readouts normalized by baseline values ($I_0$) at t=0 hr. The box plot in Fig.2B shows the average reporter signal intensity ($I_{70}/I_0$) as a function of varying exogenous TGF-β conditions under fixed IF at $\Delta P = 30$ mbar. Among all the reported conditions, we observe the strongest reporter upregulation ($I_{70}/I_0 = 13 \pm 2.73$) for an exogenous TGF-β concentration of $C_0 = 10$ ng/mL under IF (i.e. IF$^+$-TGF-β $^+$(10 ng/mL)). We also observed that supplying exogenous TGF-β (10 ng/mL) without IF (IF$^-$-TGF-β (10 ng/mL)) has approximately 73% lower reporter expression when compared to IF$^+$-TGF-β $^+$(10 ng/mL), see Fig.S8(A). This result highlights a potentiating effect of IF towards an enhanced exogenous TGF-β induced Smad-signaling activity measured via upregulation in CAGA-12-GFP transcriptional reporter response. In addition, we studied the effect of different IF velocities without exogenous TGF-β supplement for Smad-dependent CAGA-12-GFP reporter activity. We observed minimal reporter gene upregulation at t = 70 hrs which can be linked to the inactivity of the Smad-pathway in the absence of exogenous TGF-β , see Fig. S9(A and C).

To explore this potentiating effect between IF and exogenous TGF-β further, we employed time-lapse imaging to monitor the CAGA-12-GFP reporter expression profile through 70 hrs under varying IF ($u_m = 0.2$ μm/s and 0.45 μm/s at $\Delta P = 20$ and 30 mbar, respectively) and exogenous TGF-β concentrations ($C_0 = 1$ and 10 ng/mL). Fig.2C shows the time-wise variations in $\langle I/I_0 \rangle$ for the following three combinations of IF and exogenous TGF-β concentrations- IF$^+$TGF-β $^+$: $\Delta P = 30$ mbar; $C_0 = 10$ ng/ml, $\Delta P = 20$ mbar; $C_0 = 10$ ng/ml, compared with the $\Delta P = 0$ mbar; $C_0 = 10$ ng/ml i.e. no-IF condition. We observed a clear influence on the CAGA-12-GFP signal intensity profile with changing IF pressure gradients. For $C_0 = 10$ ng/ml, the IF condition at $\Delta P = 30$ mbar resulted in the fastest non-linear increase in the fluorescence signal intensity profile that begins to show saturation over the 70-hour time period (Fig.2C). For the IF at $\Delta P = 20$ mbar and the no-IF conditions, fluorescence signal intensity showed relatively slower upregulation responses (Fig.2C). Interestingly, for the IF condition of $\Delta P = 30$ mbar, decreasing the exogenous TGF-β concentration by an order of magnitude (i.e. $C_0 = 1$ ng/ml) still resulted in an upregulation response faster than the $\Delta P = 20$ mbar; $C_0 = 10$ ng/ml condition. This observation indicates that the Smad-dependent transcriptional reporter response is weakly sensitive to the exogenous TGF-β concentration, but shows a strong dependence on the IF. Under the fixed $\Delta P = 30$ mbar, for both $C_0 = 1$ and 10 ng/ml the upregulation rates are nearly equal between 25-55 hrs, after an initial delayed response for the former. Additionally, we performed similar experiments under no-IF conditions with exogenous TGF-β concentration (1 and 10 ng/mL). The CAGA-12-GFP expression for 1 ng/mL and 10 ng/mL under no-IF showed a similar upregulation profile (see Fig.S13), suggesting that the transcriptional response is fairly independent of the exogenous TGF-β concentration greater than $C_0 = 1$ ng/ml. To follow-up on the potentiating effect of IF in our 3D-microfluidic A549 spheroid experiments, we performed additional experiments in a 2D-microfluidic system (without matrix) using dual-reporter A549 cells (refer to ESI section 5 for detailed experimental procedure). In Fig. S5(C), dual-reporter A549 cells showed maximum CAGA-12-GFP reporter expression in the presence of 2D-flow and exogenous TGF-β condition. The reporter response is similar to the observations in 3D-matrix microfluidic experiments with A549 spheroids displaying maximum activity under IF and exogenous TGF-β condition (Fig. 2B). In addition, in 2D-microfluidic experiments without matrix, A549 cells showed only a modest upregulation in CAGA-12-GFP reporter expression under 2D-flow conditions without exogenous TGF-β when compared to control condition with No-flow and No TGF-β , see Fig. S5(C). This observation indicates that biophysical forces arising from 2D-flow induced fluid-shear stress feeds into the Smad-pathway that leads to CAGA-12-GFP transcriptional reporter activity. Furthermore, qPCR target gene analysis was performed on A549 cells exposed to 2D-flow and/or exogenous TGF-β conditions (refer to ESI section 6 for full experi-mental procedure) to confirm for Smad-dependent CAGA-12-GFP transcriptional reporter activity with change in TGF-β target gene expression. Maximum target gene expression for *CTGF*, *Serpin1 encoding PAI-1*, and *Smad7* was observed in A549 cells stimulated with 2D-flow and exogenous TGF-β conditions, see Fig. S6. TGF-β target gene expression for cells exposed to only 2D-flow (no exogenous TGF-β) showed an increase in *Serpin1* and *Smad7* target genes. This further highlights that flow alone activates Smad-dependent CAGA-12-GFP reporter response.

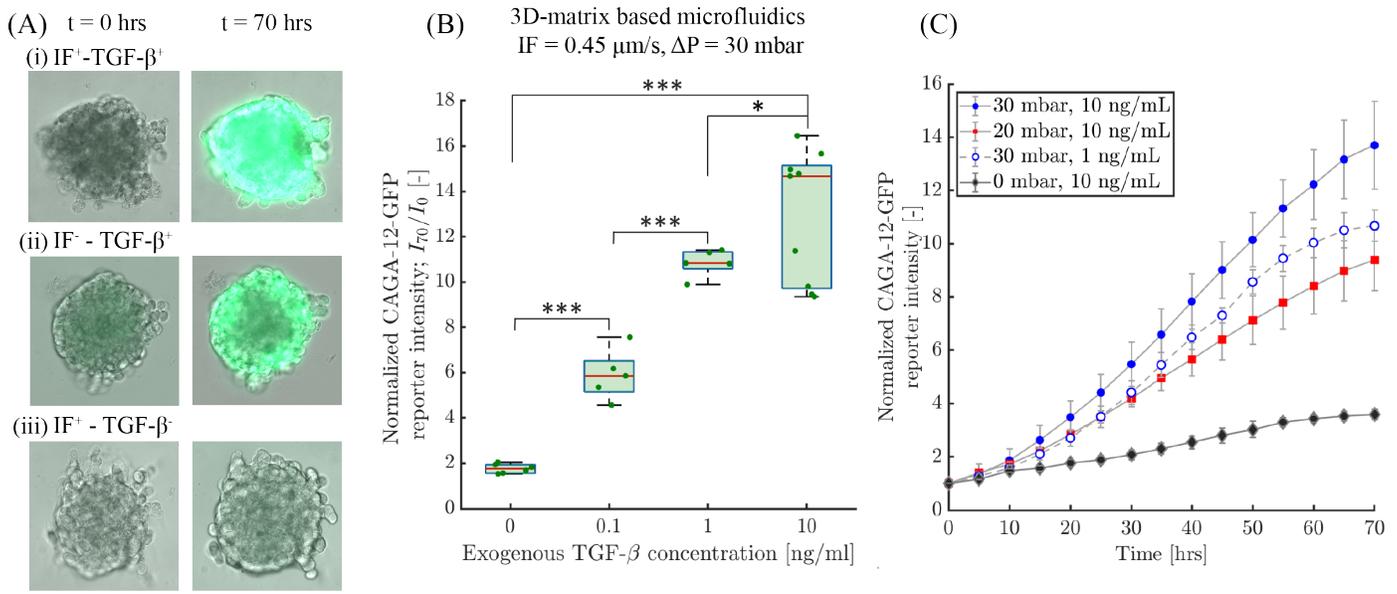

Fig. 2 Exogenous TGF-β induced CAGA-12-GFP transcriptional reporter response of A549 spheroids under interstitial flow (IF) and no-flow conditions. (A) 20x GFP and bright-field merged microscope images of A549 spheroids at t = 0 and 70 hrs showing transcriptional-reporter intensity upregulation for the following conditions: (i) IF$^+$-TGF-β $^+$(10 ng/mL), (ii) IF$^-$-TGF-β $^+$(10 ng/mL) and (iii) IF$^+$-TGF-β $^-$, scale bar: 100 μm. (B): Quantitative measurement of normalized CAGA-12-GFP reporter signal intensity at t = 70 hrs for varying exogenous TGF-β concentrations under fixed IF. (C): Time series quantification of fluorescence signal intensity (with time intervals of 5 hrs) for CAGA-12-GFP upregulation profile under different IF (ΔP = 20 mbar and 30 mbar) and exogenous TGF-β conditions (1 and 10 ng/mL), for n = 3 spheroids in each condition.

It has been proposed that the transcriptional gene response from Smad-signaling pathway is dependent on a chain of reaction kinetics initiated with binding of exogenous TGF-β molecules at the active receptor sites [45]. These reaction kinetics include expression level of TGF-β receptors and Smads and its activation state, ability to translocate into the nucleus, ability to interact with other transcription factors, co-activators, co-repressors, and chromatin modulators etc [45]. The local concentration of available TGF-β should influence the conversion capacity of available receptor sites to activated receptor sites upon successful binding. Additionally, active TGF-β availability is tightly controlled by its interaction with ECM proteins and its ability to present itself to signaling receptors is regulated by co-receptors (without intrinsic enzymatic motif), by integrins and other receptor molecules [23]. The upregulation rate of the transcriptional gene response is controlled by the density of receptor sites (i.e. number of available sites) and the reaction rate constant. To validate the TGF-β receptor binding affinity in the presence of exogenous TGF-β molecules, we used a well-known TGF-β type I receptor inhibitor (SB-431542). This small molecule inhibitor is used to inhibit all TGF-β type I receptor kinase activity. We first performed exper-iments in 3D-static (without IF) conditions with A549 spheroids embedded in 5wt.% gelMA with and without SB-431542 inhibitor treatment; followed by stimulation with exogenous TGF-β (10 ng/mL), full experimental procedure is described in ESI section 9(i). Fig. S10 shows A549 spheroids stimulated with only exogenous TGF-β showed Smad-dependent CAGA-12-GFP transcriptional activity. On the other hand, A549 spheroids initially treated with SB-431542 (10 μM) inhibitor and subsequently stimulated with exogenous TGF-β (10 ng/mL) showed no CAGA-12-GFP reporter activity. This is a result of receptor kinase activity inhibition. In addition, we saw similar effect of SB-431542 inhibitor treatment on CAGA-12-GFP and VIM-RFP reporter expression in 2D-microfluidic experiments on dual-reporter A549 single cells (without matrix) shown in Fig. S11 and S12 (full experimental procedure described in ESI section 9 (ii) and (iii)). With these experiments we concluded that, CAGA-12-GFP transcriptional reporter activity is Smad-dependent which is a result of TGF-β receptor activation upon binding with TGF-β molecules. In our 3D microfluidic studies, we estimated the evolution of local TGF-β concentration in the vicinity of a spheroid. We performed 2D mass transport simulations using the finite-element method (implemented in COMSOL Multiphysics) by varying the IF conditions and the input concentration of exogenous TGF-β (Fig. S14). At 350-400 minute mark, each condition has achieved its respective saturation concentration ($C_0$) of the exogenous TGF-β (Fig. S14). Since the upregulation of CAGA-12-GFP was found to be fairly independent of exogenous TGF-β concentration even under no-IF condition (see Fig. S13), we expect that all the active binding sites on the spheroid interface are activated for each case by this 250-400 minute mark of the experiment. Comparing this analysis with the results in Fig. 2C suggests that besides the exogenous TGF-β, there are additional biophysical forces induced mechanotransduction pathways that influence an enhanced CAGA-12-GFP reporter upregulation from TGF-β induced Smad-signaling activity.

Cancer cells have the ability to respond to mechanical cues (matrix stiffness, fluid shear stress and compressive forces) by activation of cell surface mechanosensors such as integrins, focal adhesion complex, transient receptor potential (TRP) ion channels and YAP/TAZ signaling pathway [21,46–49]. Activation of mechanotransduction signaling pathways may lead to transcriptional activity of YAP/TAZ, commonly identified to promote cancer cell invasion and trigger EMT signaling pathways [20,50,51]. Earlier studies have linked mechanotransduction induced EMT for cancer cells under a flow-induced shear stress of 0.1 - 3 Pa [52–54]. These studies were performed on 2D monolayer culture without an extracellular matrix environment. The shear stress induced by an interstitial fluid flow is typically reported to be in the order of 0.01 Pa [55]. These values are reported for cells cultured on 2D substrate subjected to interstitial flow velocities in microfluidic systems. In our 3D-matrix based microfluidic study, we find that IF generated via a pressure gradient leads to both flow-induced shear stress and compressive stress contributing to the Smad-dependent transcriptional reporter activity. Based on our simulation, the shear stress on a 2D spheroid model interface embedded in a low permeability matrix (mimicking the properties of gelMA used in the experiments) was found to be relatively low (∼ 0.1-0.3 mPa, see Fig. S3(C)). The compressive/normal stress caused by hydrodynamic pressure at the spheroid interface is significantly high (∼ 1 and 2 kPa, see Fig. S3(D)). Previous literature has highlighted the role of matrix stiffness and matrix-induced compressive forces activating key mechanotransduction pathways (Wnt, Hippo, PI3-AKT, TGF-β) for cancer cell proliferation and migration [56–58]. Since our 3D-microfluidic platform does not allow access to A549 spheroids embedded in hydrogel matrix, we are unable to perform qPCR analyses to confirm for specific mechanotransduction pathway underplay. A detailed study to identify specific mechanotransduction pathways under IF upregulated in a 3D-TME exposed to IF will shed more light on this hypothesis.

### 3.2 Local fluorescence profile of CAGA-12-GFP reporter activity in A549 spheroid under varying IF-exogenous TGF-β condition

To examine the local fluorescence profile of Smad-dependent CAGA-12-GFP transcriptional reporter activity in a spheroid as a consequence of varying IF from different pressure gradient, we compared evolution of fluorescence intensity at different times for a fixed exogenous TGF-β concentration ($C_0$ = 10 ng/mL) under two different values of IF (ΔP = 20 and 30 mbar) and no-IF conditions. To represent the local heterogeneity in fluorescence intensity of a spheroid exposed to varying IF, we used the Polar Transformer function in ImageJ. An example of methodology for this analysis technique on one set of spheroid images at different time intervals is shown in ESI section 12. This image analysis function converts a 2D-microscope image from a Cartesian coordinate system to a Polar coordinate system (r, θ), see Fig. S15 (A and B). Using this function, we then measured the radially averaged intensity of a polar coordinate (r, θ) transformed image at different time intervals (t = 0, 24, 48 and 70 hrs) of a particular spheroid, see Fig. S15(B). We then plot the radially averaged intensity on the azimuthal scale, i.e. θ = 0 to 360 degrees, see Fig. S15(C), which constructs the evolution of fluorescence intensity profile corresponding to the spheroid fluorescence intensity at different time intervals. The intensity profile for each spheroid was normalized to its initial fluorescence value at t = 0 hrs. The difference in intensity of fluorescence signal among these conditions are influenced by the varying IF conditions (as previously discussed in section 3.1, Fig.2C). Fig.3 compares the intensity profiles of spheroids under varying IF (Fig.3A and B) and no-IF condition (Fig.3C). The fluorescence intensity ($I(t)/I_0$) is plotted on the scale 0-15 (represented in blue). Averaged fluorescence intensity at each time point (denoted with different colors) is represented with a solid line and standard deviation in intensity with its corresponding shaded region. The azimuthal axis, θ (counterclockwise, in red) is used to represent the local fluorescence profile of the spheroid. When spheroids are exposed to IF, we can observe the asymmetry by the averaged fluorescence intensity profile of spheroids (n = 3) at θ = 90 (top) and 270 (bottom) degrees, shown in Fig.3. In Fig.3A, spheroids under IF at ΔP = 30 mbar, show an average fluorescence intensity at the top of the spheroid (θ = 90 degree) is 12.3 ± 2.65 and at the bottom (θ = 270 degree) is 9.8 ± 2.17 at t = 70 hrs (in purple). We observe a similar trend in heterogeneity of fluorescence intensity for spheroids under IF at ΔP = 20 mbar (Fig.3B). Fig.3C shows that the fluorescence intensity profiles of spheroids under no-IF condition shows axisymmetry, i.e. no noticeable change in fluorescence intensity at the top/bottom of the spheroid. In no-IF condition, fluorescence intensity is not influenced by any hydrodynamic effect or fluid induced shear/compressive stress. We suspect that the absence of biomechanical stress (IF) and inactivation of mechanotransduction pathways justifies axisymmetric fluorescence profiles in only exogenous TGF-β exposed A549 spheroids. The top-bottom asymmetry in CAGA-12-GFP upregulation profiles along the direction of flow (top to bottom) is a result of the applied IF conditions originating from varying hydrodynamic pressure.

### 3.3 Exogenous TGF-β induced vimentin activity towards cancer cell motility exposed to interstitial flow in A549 spheroids

To further explore the potentiating effect of interstitial flow (IF) with exogenous TGF-β, we examined the upregulation of vimentin as measured by determining the VIM-RFP reporter response. Vimentin, a key mesenchymal biomarker, is upregulated in lung cancer cells in the presence of TGF-β towards EMT response [25,27,28]. We measured the upregulation in vimentin expression activity by quantifying VIM-RFP reporter expression under IF and no-flow conditions in the presence of exogenous TGF-β. Fig.4A shows the superposed microscope images of brightfield and RFP channels at t = 0 and 70 hrs. We observed an enhanced VIM-RFP reporter expression with IF and exogenous TGF-β (IF$^+$TGF-β $^+$) (Fig. 4A (i)) compared to IF$^-$TGF-β $^+$ (Fig. 4A (ii)) condition). Fig.4B shows the quantified VIM-RFP signal upregulation for the same conditions (described in section 3.1). We observed that the strongest reporter upregulation ($I_{70}/I_0$ = 3.7± 0.74) with an exogenous TGF-β concentration of 10 ng/mL under IF (∆P = 30mbar) (i.e. IF$^+$-TGF-β $^+$(10 ng/mL)). We also observed that the upregulation of VIM-RFP intensities for TGF-β concentration 10 ng/mL and 1 ng/mL ($I_{70}/I_0$ = 3.6± 0.2) under IF (Fig.2B) showed no significant difference. Moreover, the VIM-RFP expression induced by exogenous TGF-β (10 ng/mL) under no-flow ($I_{70}/I_0$ = 1.41± 0.08) is 62% lower compared to spheroids under IF with exogenous TGF-β (10 ng/mL), refer Fig. S8(B). In addition, we studied the effect of varying IF without exogenous TGF-β for VIM-RFP reporter response. We observed no change in reporter activity at t = 70 hrs in A549 spheroids as a result of varying IF conditions, see Fig. S9 (B and C). Following up with 2D-microfluidic experiments, we observed maximum VIM-RFP expression for 2D-flow and exogenous TGF-β conditions, similar to observations made in 3D-microfluidics in A549 spheroids. In addition, A549 cells exposed to 2D-flow (without matrix) showed upregulation in VIM-RFP reporte expression when compared to No-flow No TGF-β (control) condition, see Fig. S5(B and D). Therefore, we propose that the IF in the presence of exogenous TGF-β has a potentiating effect that is further responsible for producing an increased VIM-RFP reporter expression corresponding to a higher vimentin abundance. These results highlight the potential involvement of mechanotransduction induced signaling pathways that contribute towards upregulation in mesenchymal marker in A549 cells.

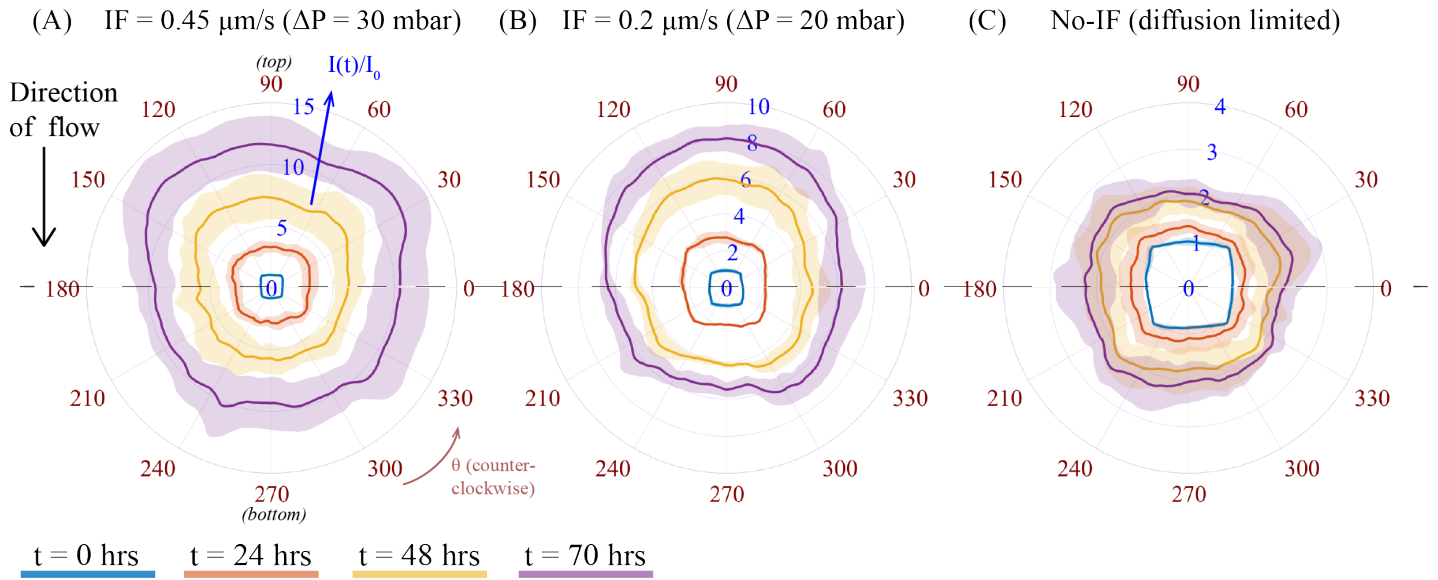

Fig. 3 CAGA-12-GFP reporter fluorescence profile under a fixed exogenous TGF-β concentration (10 ng/mL) and varying interstitial flow and no-flow conditions. Polar plot of radially averaged fluorescence intensity is denoted in $I(t)/I_0$ (in blue) at different time intervals (represented with solid lines of different color). The solid line and the shaded region of a particular color of a particular time point show averaged and standard deviation of fluorescence intensity respectively for n = 3 spheroids. The evolution of fluorescence intensity at the top and bottom half of the spheroid (from the dashed line) shows the CAGA-12-GFP reporter expression profile for conditions: (A) IF$^+$(0.45 µm/s)TGF-β $^+$, (B) IF$^+$(0.2 µm/s)TGF-β $^+$, and (C) No flow, IF$^-$TGF-β $^+$.

We quantified spheroid peripheral activity for increased cellular motion activity. The cellular motion activity here is referred to as cells at the edge of a spheroid that responds to biophysical and biochemical cues characterized with an increased motility. When stimulated with exogenous TGF-β and IF, we observed an increase VIM-RFP reporter expression corresponding to increase in vimentin abundance protein. This is identified as a mesenchymal biomarker, that is often associated with a phenotype observed in motile cancer cells. To quantify cellular motion activity, we performed temporal standard-deviation analysis of bright-field time-lapse images. This analysis technique detected the change in pixel intensity value at each time point for the duration of the entire experiment. After processing all images, the final image represents the qualitative measurement of the standard deviation change in pixel value corresponding to the cellular motion activity at the spheroid periphery. Fig.4C and D demonstrates a comparison of an A549 spheroid under IF (0.45 µm/s, ∆P = 30 mbar) and no-IF both stimulated with exogenous TGF-β (10 ng/mL). Refer to supplementary movie S1 and S2 (corresponding to spheroid in Fig.2C and D respectively) for time lapse video of cellular motion at spheroid edges embedded in gelMA matrix. In Fig.4C, the A549 lung tumor spheroid shows increased cellular motion activity under IF$^+$-TGF-β $^+$ (10 ng/mL) condition with a larger standard deviation measured correlating with increased cellular motion activity. The increase in cell cellular motion activity was mostly observed at the top/side section of the spheroid periphery. From Fig.4D, only exogenous TGF-β is insufficient to produce cellular motion activity depicted with low standard deviation of pixel value change. In these conditions, the A549 spheroid periphery does not show active cellular motion (refer to Supplementary movie, S2). The increased activity in the presence of IF and exogenous TGF-β condition can be linked to the hypothesis of mechanotransduction pathway induced activity. Additional experiments performed in 2D-microfluidics for qPCR analyses (refer to ESI section 6, Fig.S7) showed that dual-reporter A549 cells exposed to 2D-flow (without matrix) led to downregulation in *E-cadherin* expression and upregulation in *N-cadherin* and *Vimentin* characteristic of EMT target genes. This brings us closer to our hypothesis of normal/shear stress activating mechanosensors for triggering additional mechanotransduction signaling pathways. These findings highlight the importance of IF and exogenous TGF-$\beta$ that directly influence A549 tumor cells to undergo active cellular motion in a tumor microenvironment.

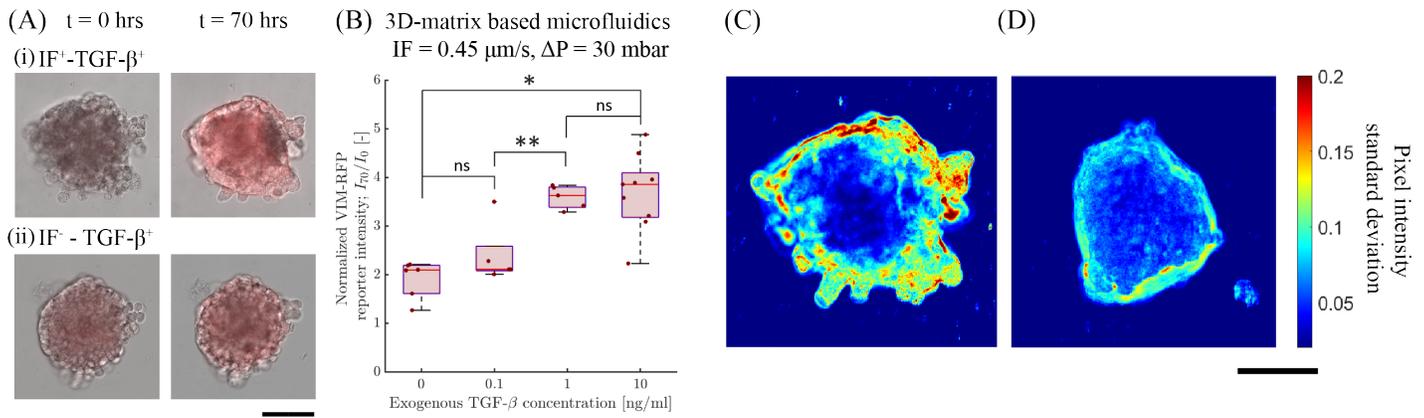

Fig. 4 Upregulation in vimentin expression as measured via VIM-RFP reporter gene response toward cell motility in A549 spheroids. (A) 20x RFP and bright-field channel merged microscope images of A549 spheroids at t = 0 and 70hrs showing gene-reporter intensity upregulation in the following conditions. (i) IF$^+$-TGF-$\beta$ $^+$(10 ng/mL) and (ii) IF$^-$-TGF-$\beta$ $^+$(10 ng/mL), scale bar: 100 $\mu$m. (B) Quantitative measurement of normalized VIM-RFP reporter signal intensity at t = 70 hrs for varying exogenous TGF-$\beta$ under fixed IF (0.45 $\mu$m/s) at $\Delta$P = 30 mbar. (C) and (D) Standard deviation analysis showing cellular motion activity at spheroid periphery of A549 spheroids under IF and no-flow conditions with exogenous TGF-$\beta$ (10 ng/mL), scale bar: 100 $\mu$m.

## 4 Conclusions and Outlook

We used a 3D-matrix based microfluidic platform to investigate the potentiating effect of IF on exogenous TGF-$\beta$ induced Smad-signaling activity in A549 lung cancer spheroids. Our platform allowed us to embed cancer spheroids in 3D using gelMA hydrogel as a relevant ECM material. This integrated platform of porous hydrogel material and cancer spheroid allowed us to mimic interstitial flow (IF) conditions experienced by a tumor in a TME. One advantage of this microfluidic platform was the ability to investigate cancer cell-matrix interactions over time, allowing us to observe the effects of varying biophysical conditions and biochemical signals. By studying the interplay between biophysical components (hydrogel matrix and IF), and the externally introduced cytokine (exogenous TGF-$\beta$), we aimed to better understand how these factors contribute to cancer spheroid response and invasive behavior. To this end, we monitored the upregulation in transcriptional reporter response (CAGA-12-GFP) and vimentin abundance protein (VIM-RFP reporter) in A549 lung spheroids using real-time imaging of artificial gene reporter constructs. Our findings suggest that the addition of IF within the 3D-matrix significantly enhances the CAGA-12-GFP reporter response from Smad-signaling activities upregulated by exogenous TGF-$\beta$. This also leads to an increase in the abundance of vimentin protein measured via upregulation in VIM-RFP reporter expression and increased cellular motion activity observed at the spheroid periphery in a matrix microenvironment. Additional experiments performed in 2D-microfluidic system (without matrix) further confirmed the mRNA expression of TGF-$\beta$ and EMT target genes to establish CAGA-12-GFP and VIM-RFP reporter functionality of the dual-reporter A549 cells. In 2D-microfluidic experiments (without matrix), flow (without exogenous TGF-$\beta$) alone has a clear evidence of upregulation in Smad-dependent CAGA-12-GFP reporter activity and *vimentin* gene with qPCR analyses. Moreover, 2D-flow also resulted in downregulation of *E-cadherin*. The fluorescent reporter activity confirmed by target gene analyses indicates that 2D-flow alone activates parallel pathways that feed into the Smad pathway leading to Smad-induced transcriptional reporter activity. In 2D-microfluidic (without matrix) and 3D-microfluidic (with matrix), CAGA-12-GFP and VIM-RFP showed maximum reporter intensity when exposed to flow and exogenous TGF-$\beta$. However, in 3D-microfluidic experiments, A549 spheroids embedded in gelMA showed no clear upregulation in reporter expression when exposed to only IF (without exogenous TGF-$\beta$), in contrast to 2D-microfluidic experiments. The reporter expression was potentiated by IF only in the presence of exogenous TGF-$\beta$. The reporter expression showed clear dependency on the magnitude of IF applied (increasing pressure gradient) and showed less sensitivity to exogenous TGF-$\beta$ concentration. Using complimentary numerical simulation on a 2D spheroid model, we further characterized the mass transport of TGF-$\beta$, flow induced shear and normal stresses on the spheroid interface under different IF conditions. Based on these results and qPCR analyses, we hypothesize that exogenous TGF-$\beta$ induced Smad-signaling and vimentin expression is further upregulated from potentiating effect of interstitial flow mediated mechanotransduction pathways.

The 3D microfluidic platform introduced in this study has the potential to expand beyond tumor spheroid models, and can be applied to heterogeneous tumor spheroid, stromal cells, cancer-associated fibroblasts (CAFs) and immune cells. This versatility brings us closer to mimicking in-vivo tumor microenvironment

(TME) conditions.

## Author Contributions

## Conflicts of interest

There are no conflicts to declare.

## Acknowledgements

Z.R., A.B. and P.E.B gratefully acknowledge funding from the European Research Council (ERC) under the European Union's Horizon 2020 research and innovation program (grant agreement no. 819424). P.T.D. and P.E.B. gratefully acknowledge funding from the Delft Health Technology grant (between LUMC and TU Delft) and ZonMW grant (09120012010061). A.B. gratefully acknowledges funding from MSCA Postdoctoral Fellowships 2022 Project ID: 101111247. The authors thank Yifan Zhu for the dual-reporter A549 cells.

# Interstitial flow potentiates TGF-β/Smad-signaling activity in lung cancer spheroids in a 3D-microfluidic chip – Supplementary Material

Zaid Rahman,[a] Ankur Bordoloi,[a] Haifa Rouhana,[a] Margherita Tavasso,[a] Gerard van der Zon,[a] Valeria Garbin,[a] Peter ten Dijke,[b] and Pouyan E. Boukany [*a]

[a] Department of Chemical Engineering, Delft University of Technology, Delft, The Netherlands.
[b] Department of Cell and Chemical Biology and Oncode Institute, Leiden University Medical Center, Leiden, The Netherlands
[*] corresponding author

## Electronic Supplementary Information (ESI)

### 1. Storage Modulus and stiffness characterization

The viscoelastic properties of GelMA were investigated with a modular rotational rheometer (DSR 502, Anton Paar) equipped with a parallel plate of diameter of 25 mm. A volume of 500 µL of gel was crosslinked on a glass slide in the same conditions as in the microfluidic device. Amplitude and frequency sweeps allowed to estimate the strain and time-dependent behavior of the sample. Frequency sweeps were performed from low to high frequencies and vice-versa (0.1 to 100 rad/s at a fixed strain of 1%) at room temperature (22$^0$C). The results are similar as the measurements occur within the linear viscoelastic (LVE) regime of the gel.

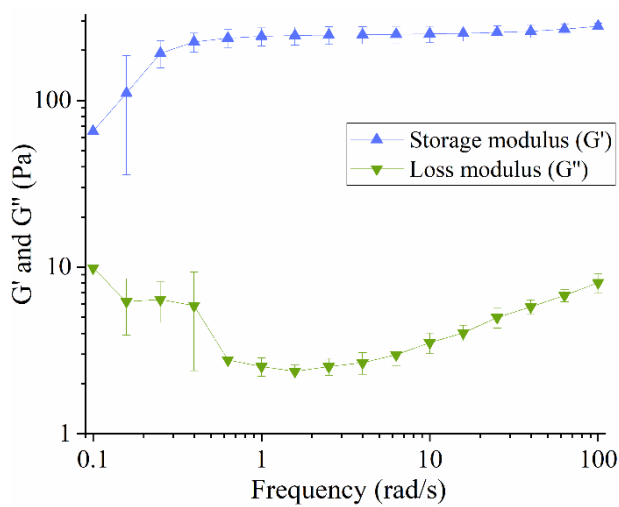

**Figure S1: Frequency dependence of storage modulus (G') and loss modulus (G'') of 5 wt.% gelMA:** Characterized in small-amplitude oscillatory shear analysis (at a fixed strain amplitude of 1.0% at room temperature).

### 2. Design and fabrication of 3-D microfluidic chip

The master wafer was fabricated by a standard photo-lithography technique using a µMLA laserwriter (Heidelberg Instruments) on a 4-inch silicon wafer at the Kavli Nanolab Delft. The top and the bottom channel are fluidic channels and the middle channel is separated via triangular shaped pillars where spheroids are suspended in a hydrogel. The hydrogel is confined in the middle channel primarily due to the presence of evenly spaced triangular posts[1,2]. We optimized the photolithography procedure to achieve a final channel height of 280 ± 25 µm to embed cancer spheroids in 3D. To obtain this, we first spin-coated SU-8 2150 (acquired from Kayaku Advanced Materials) negative photoresist in two steps. First step was at 500 rpm for 10 seconds with an acceleration of 100 rpm per second followed by 2300 rpm for 30 seconds with an acceleration of 300 rpm per second. The silicon-wafer with SU-8 was then transferred to a hot plate to soft-bake at $65^0$C for 15 minutes, followed by $95^0$C for 90 minutes. After soft-bake, the wafer was loaded onto the laserwriter sample holder stage. The AutoCAD design was uploaded to the laserwriter software to print the design using a 365 nm laser source. The wafer was post baked at $65^0$C for 15 minutes, followed by $95^0$C for 5 minutes and developed in SU- 8 photo-resist developer (mr-Dev 600, Micro resist technology). The microfeatures are constantly checked under an optical microscope to adjust the photo-resist developing time period in order to avoid over or under-development. The height of the channels was determined by a Dektak profilometer. The profilometer measures channel height 280 µm ± 25 µm. The master wafer was coated with trichloro(1H,1H,2H,2H-perfluorooctyl)silane to create a hydrophobic surface for easy demolding. Polydimethylsiloxane (PDMS) based microfluidic chips were prepared using the mixture of Sylgard 184 and curing agent (at a ratio of 10:1). Individual chips were then cut out and inlets/outlets were punched in the PDMS slab consisting of the design to insert tubings (PTFE, inner diameter: 0.8mm, outer diameter: 1.6mm) for fluidic-channel operation. Final bonding is performed by plasma cleaning (Harrick Plasma) of the PDMS slab and a glass coverslip (#1.5) for two 2 minutes 20 seconds to facilitate final bonding. Finally, the assembled device was kept in the oven at $65^0$C to bond overnight and restore the hydrophobicity of the channels. Afterwards, the microfluidic devices can be stored indefinitely and sterilized by UV-irradiation for 20 minutes before the start of each experiment.

## 3. Hydraulic permeability (K) measurement to compute interstitial flow velocity

To quantify interstitial flow velocity through 5 wt.% gelMA hydrogel, we first identified the hydraulic permeability (K) of the hydrogel. We used Rhodamine B (fluorescent dye) as tracer molecule and measured the distance covered by the fluorescent dye through gelMA in a specific time frame. These studies are performed only with the hydrogel material and no spheroids. The microfluidic chip was prepared by first filling the hydrogel channel with 5wt.% gelMA, followed by crosslinking using 5x-obective lens with a LED UV-light source (385 nm) from Colibri Inverted microscope for 45 seconds. We prepared the dye solution by mixing a 1% (v/v) solution of Rhodamine B in 1X Dulbecco's Phosphate Buffer Solution (DPBS, Sigma Aldrich). This dye solution is used as a fluid reservoir connected to the Fluigent MFCS-EZ pressure pump to allow flow into the microfluidic channels. The tubings (inner diameter: 0.8 mm and outer diameter: 1.6 mm) from the reservoir are connected to the inlets ($P_1$) of the top channel. The bottom channel inlets ($P_2$) were connected to an empty reservoir. By applying a pressure gradient, we generated a flow through the porous hydrogel material, see Fig. S2 (for experimental setup). To validate IF, we tracked fluorescent dye travel with time-lapse imaging at an interval of 30 seconds. Images were taken on Zeiss Axio Observer Colibri 7 equipped with ORCA Flash 4.0 V2 (Hamamatsu) digital camera with a resolution of 2048 x 2048 pixels using 543nm LED laser source (excitation/emission: 543 nm/568nm), at 30% intensity laser and 1.58 seconds exposure time, with objective lens set to 5x. Flow velocity was then calculated using ImageJ for pressure gradients of ΔP: 20 mbar and 30 mbar. These pressure gradients were optimal since a high pressure gradient (> Δ40 mbar) disturbs the mechanical stability of the hydrogel and disrupts uniform interstitial flow. Images are exported from Zeiss software as .tiffiles and imported as a stack into ImageJ for analysis. The images were then straightened and converted into 8-bit color images. A vertical line was drawn as area of interest to plot profile of gray value vs. distance. Mean gray value of zero corresponds to the last position of the dye at a specific time, this position was recorded for consecutive images for an average of 10 time points. The velocity was then calculated by dividing the distance traveled by the duration (between two consecutive images) and the average velocity on the line of interest is obtained. . The average velocity for ΔP = 20 mBar and 30 mBar is calculated at 0. 46 ± 0.21 µm/s and 0.72 ± 0.16 µm/s respectively (see Table 1 for average velocity measurements). All known parameters were plugged into Darcy's law (equation 1) to obtain the hydraulic permeability (K) of the hydrogel.

$$v = \frac{\Delta P \times K}{L \times \mu} \quad (1)$$

Where $v$ is the fluid velocity across the porous network (m/s), $\Delta P$ is the pressure gradient across the porous network (Pa), $L$ is the length of the porous network in the direction of the flow (m), $K$ is the hydraulic permeability (m²), and $\mu$ is the fluid dynamic viscosity (Pa s). The average permeability constants were measured at approximately 1.35x $10^{-16}$ m² for 5wt.% gelMA. The permeability of 5wt.% gelMA was compared to the values found in literature to be of approximately 0.1 µm²/Pa.s for a high degree substitution 5wt.% gelMA[3], equivalent to $10^{-16}$ m².

| Time (sec) | ΔP = 30 mbar | | ΔP = 20 mbar | |
|---|---|---|---|---|
| | Distance travelled (µm) | Calculated average velocity (µm/s) | Distance travelled (µm) | Calculated average velocity (µm/s) |
| 0 | 0 | 0 | | |
| 30 | 23.97 | 0.79 | 8.20 | 0.27 |
| 60 | 32.18 | 1.07 | 21.54 | 0.71 |
| 90 | 24.61 | 0.82 | 18.29 | 0.61 |
| 120 | 18.23 | 0.60 | 10.09 | 0.33 |
| 150 | 15.14 | 0.50 | 22.71 | 0.75 |
| 180 | 22.72 | 0.75 | 5.04 | 0.16 |
| 210 | 22.08 | 0.73 | 12.62 | 0.42 |
| 240 | 22.08 | 0.73 | 20.82 | 0.69 |
| 270 | 14.51 | 0.48 | 14.51 | 0.48 |
| 300 | 23.35 | 0.77 | 4.41 | 0.14 |
| | Average velocity | 0.72 | Average velocity | 0.45 |
| | Standard deviation | 0.16 | Standard deviation | 0.26 |

Table S1: Average velocity measurement for ΔP = 20 and 30 mbar.

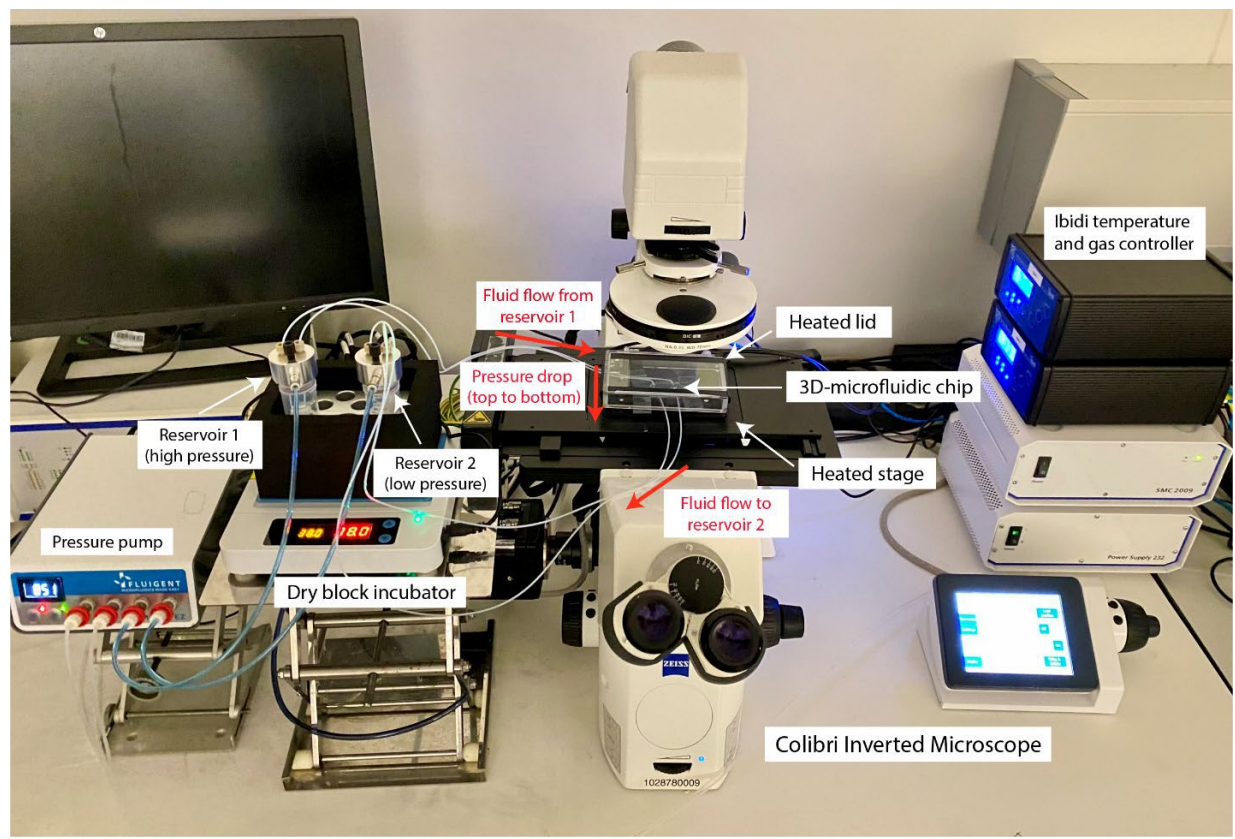

**Figure S2: Experimental setup to generate interstitial flow in a 3D-microfluidic chip using a pressure pump.** The reservoirs are kept in a dry block incubator maintained at 37⁰C. The pressure pump channels are connected to the media reservoirs set at high and low pressure values to create a pressure gradient. The microfluidic chip is placed in a Ibidi stage top incubator equipped with a heated lid and a heated stage. The Ibidi temperature and gas controller box is used to maintain cell culture conditions at $37^{0}$C, 95% humidity and 5% $CO_2$. Colibri inverted microscope used for time series fluorescence imaging.

## 4. Fluid velocity and stress profiles around the spheroid

To compute the interstitial flow velocity in the microfluidic model, we solved 2D incompressible Stoke's flow equations using inbuilt Free and Porous Media Flow interface in COMSOL Multiphysics. For simplicity, the spheroid was modeled as a solid circular region with a rigid boundary. A pressure gradient across the porous matrix was generated by imposing pressure $P_1$ at the inlets (A and B) and $P_2$ ($<P_1$) at the outlets (C and D). The velocity and pressure fields were computed over the domain discretized using a physics based unstructured mesh with approximately $3\times10^5$ elements with a minimum size of 1 μm, represented in Fig.S3(A). We modeled the hydrogel material using the permeability value (K) of 5wt.% gelMA calculated in section 2. Fig. S3(B) shows a representative velocity field around a spheroid model for ΔP = 30 mbar. The computed average flow velocities for ΔP = 20 and 30 mbar are 0.2 μm/s and 0.45 μm/s, respectively. Fig. S3(C) shows the statistics of local shear stress at the spheroid interface for the two imposed pressure gradients. The maximum fluid-induced shear stress was measured at 0.3 mPa at the spheroid interface for ΔP = 30 mbar condition. Fig. S3(D) and S3(E) show the normal stress measured at ~1750 Pa at the top spheroid interface for ΔP = 30 mbar condition and ~1250 Pa for ΔP = 20 mbar condition, respectively.

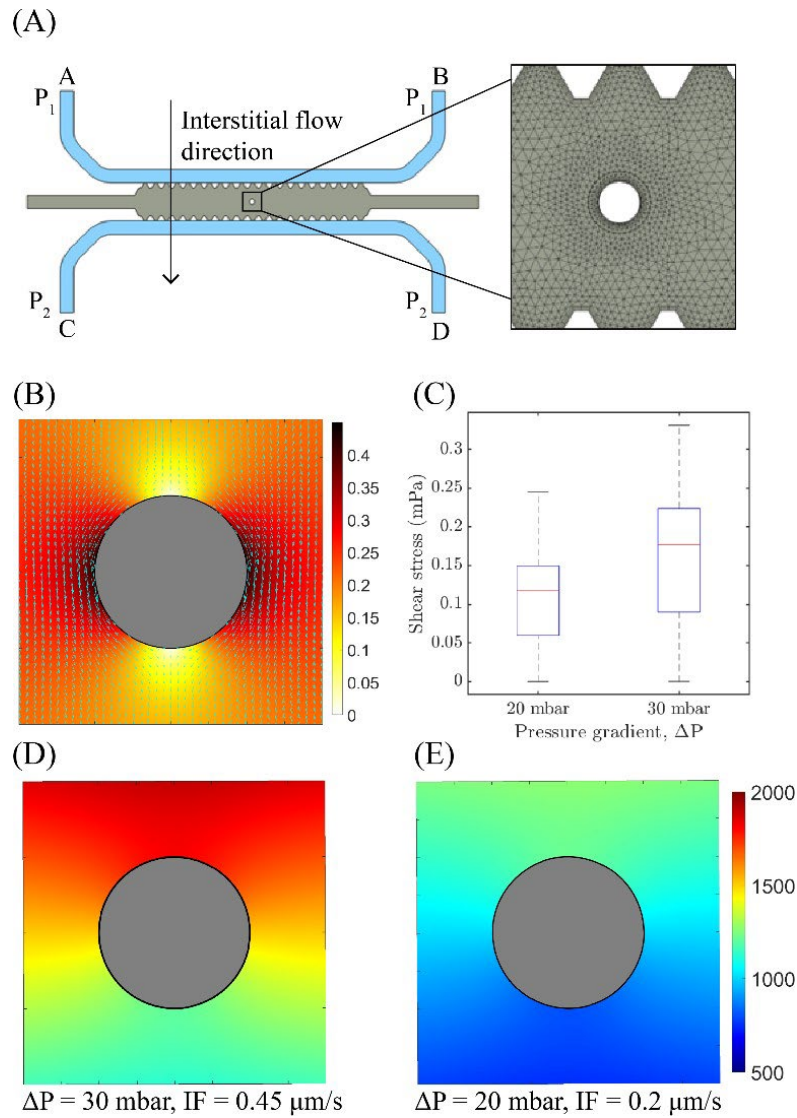

**Figure S3: COMSOL simulation to compute interstitial fluid velocity and flow induced shear and normal stresses** (A) 2D microfluidic model used for performing numerical simulation and the typical unstructured distribution near the spheroid model. (B) An example demonstrating the velocity field (absolute velocity) around the spheroid model at ΔP = 30 mbar (direction of flow: top to bottom, scale: μm/s). (C) IF-induced shear stress distribution at the spheroid interface under two pressure gradient conditions. (D,E) Normal stress experienced by the spheroid model at ΔP = 30 mbar and 20 mbar respectively (scale: Pa).

## 5. 2D microfluidic flow experiments with dual-reporter A549 cells for qPCR analyses

To further investigate the effect of flow conditions on the upregulation of Smad-signalling pathway and mesenchymal marker vimentin, we performed experiments with dual-reporter A549 cells in a 2D-microfluidic device (without matrix) subjected to flow with and without the presence of exogenous TGF-β. The 2D-flow experiments are performed in a microfluidic chip with a microchannel of dimensions 2 cm x 1 mm x 0.1 mm (length x width x height) as shown in Fig. S4, fabricated using standard soft-lithography technique, described in ESI section 2. The microchannels were coated with 20 μg/mL fibronectin (Sigma

Aldrich, St. Louis, MO, USA) and kept in a humidified incubator for 6 hrs before seeding dual-reporter A59 cells at a density of $10^5$ cells/mL. The microfluidic chip with seeded cells were then kept in a humidified incubator for overnight attachment to achieve a confluent 2D-monolayer. After successful cell attachment, we imaged the cells for their reporter expression (CAGA-GFP and VIM-RFP) at t = 0 hrs with a Zeiss LSM 980 Confocal Microscope at 20x/NA 0.8 M27 air objective at 1024 x 1024 pixel$^2$ density. For GFP and RFP fluorescence, lasers with excitation wavelength of 488 nm and 543 nm were used, respectively. For the experiment, we prepared two reservoirs, one containing cell culture media supplemented with exogenous TGF-β and the other without exogenous TGF-β. To clearly identify the effect of 2D-flow, two different microfluidic devices were used, one connected to reservoir with TGF-β and the other without TGF-β. The reservoirs and microfluidic chips were placed in a humidified incubator. Next, we connected two syringes on a syringe pump (Harvard Apparatus, Elite 2000) placed outside the incubator. The tubings from the syringes were connected to the outlets of the microfluidic devices inside the incubator. The flow rate was set at 0.1 µL/min using the withdrawal function. We calculated the fluid-induced wall shear stress from the formula;

$$\tau = \frac{6 \times \eta \times Q}{W \times h^2} \qquad (2)$$

Where η is the viscosity (in Pa.S), Q is the flow rate (m$^3$/s), W is the width (m) of the channel and h is the height (m) of the microfluidic channel. This flow rate creates a fluid-induced wall shear stress at $1 \times 10^{-3}$ Pa. The experiment was performed for 72 hrs and imaged using the same image settings described above. Simultaneously, we prepared samples under 2D-static (no-flow) conditions where A549 cells were stimulated with and without exogenous TGF-β, imaged at t = 0 and t = 72 hrs. Fig. S5 (A and B) shows the merged GFP + BF and RFP + BF microscope image of reporter expression CAGA-GFP and VIM-RFP respectively of 2D cultured A549 cells at t = 72 hrs for the following conditions: i) No-flow No TGF-β, ii) No-flow + TGF-β, iii) Flow – No TGF-β, iv) Flow + TGF-β. We analyzed the upregulation in fluorescence intensity for all cases and plotted the intensity measured at t = 72 hrs normalized with respect to control sample at t = 72hrs, see Fig. S5(C and D). The reporter expression in 2D-flow (without matrix) condition revealed the activity of Smad-induced CAGA-12-GFP reporter response as well as VIM-RFP indicating abundance of vimentin protein in A549 cells. The experiment was performed in duplicates and each experiment was imaged at 3 different locations of the microfluidic channel (total n = 6 locations for each condition in 2 independent experiments). Statistical analysis on each set was performed using Student's t-test (paired two sample).

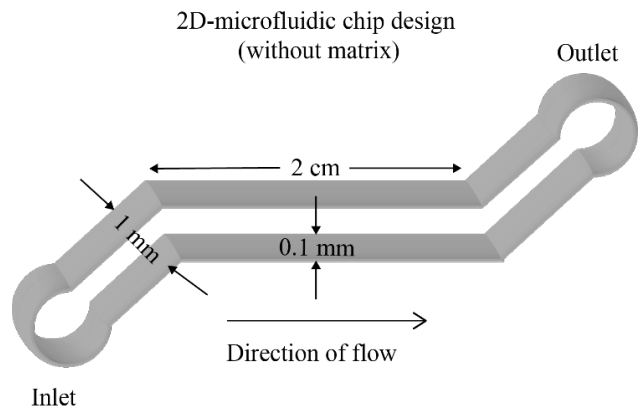

**Figure S4: 2D-microfluidic chip design (with dimensions, not to scale) for experiments performed with dual reporter A549 single cells under 2D-flow and exogenous TGF-β conditions without hydrogel matrix.**

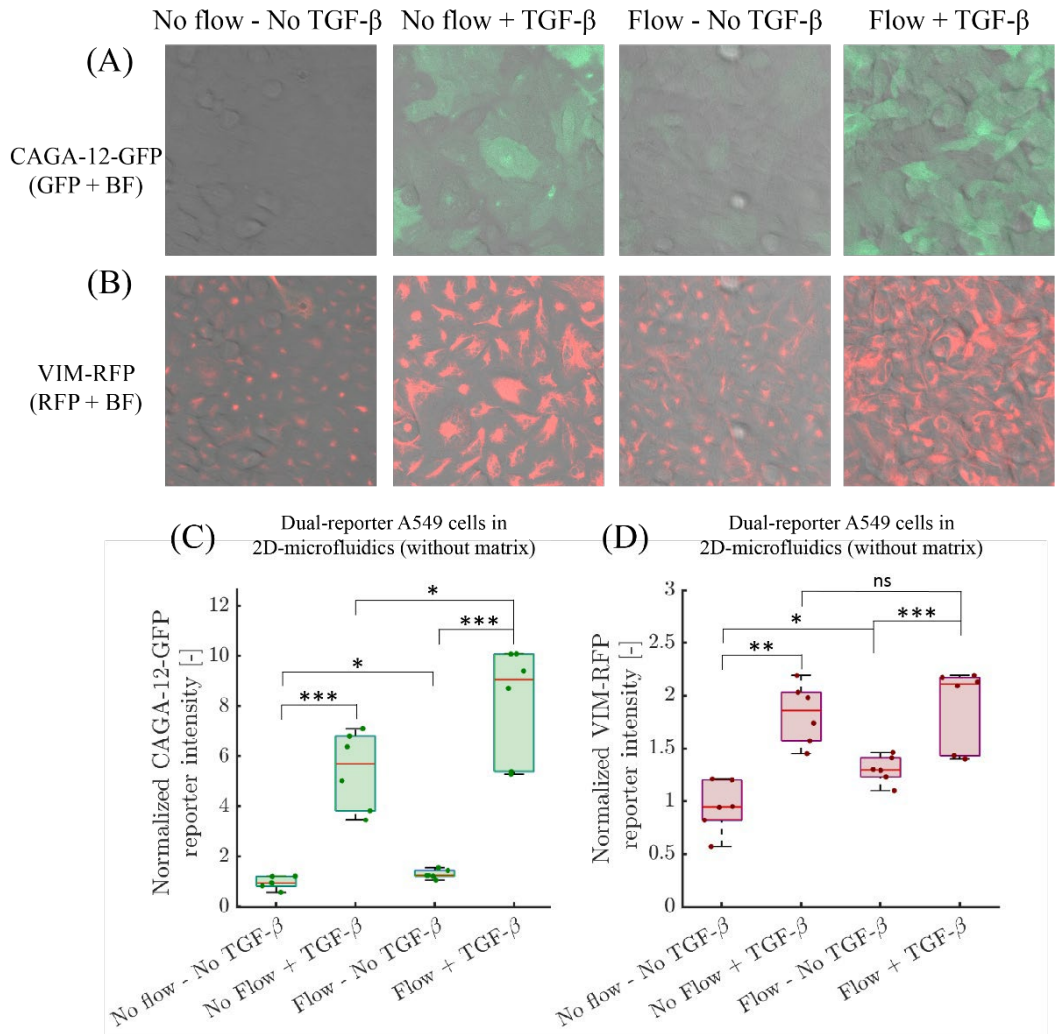

**Figure S5: 2D-microfluidic (without matrix) experiments performed with dual-reporter (CAGA-12-GFP and VIM-RFP) A549 single cells.** (A) Bright-field and GFP channel (merged) image at t = 72 hrs for all experimental conditions for CAGA-12-GFP reporter expression of A549 cells. (B) Bright-field and RFP channels (merged) imaged at t = 72 hrs for VIM-RFP reporter expression of A549 cells. (C) Quantification of CAGA-12-GFP fluorescence signal intensity at t = 72 hrs, normalized to fluorescence intensity of No flow – No TGF-β (control) condition at t = 72 hrs in 2D-microfluidics. (D) Quantification of VIM-RFP fluorescence signal intensity at t = 72 hrs, normalized to No flow – No TGF-β (control) condition at t = 72 hrs in 2D.

## 6. RNA isolation, cDNA synthesis and real time-quantitative PCR (RT-qPCR)

After the experiment, the cells from the microchannel are first extracted using trypsin for 3 min and diluted in fresh culture medium followed by centrifugation at 200g for 5 mins. The supernatant is pipetted out leaving behind the cell pellet. Total RNA extraction is performed using NucleoSpin RNA II kit (Macherey-Nagel, Deuren, Germany) according to the manufacturer's instructions. cDNA was generated using the Revert Aid First-Strand cDNA synthesis mix (Thermo Fisher, Bleiswijk, The Netherlands). Quantitative PCR was performed using SYBBR GoTaq qPCR master mix (Promega, Leiden, Netherlands) using 0.5 µM of primers. RT-qPCR was performed on the CFZ connect Real-Time PCR detection system (Bio-Rad, Hercules, CA, USA). The primers used are listed in Table S2. All experiments were performed in triplicates, target gene expression was normalized to the geometric mean (geomean) of *ARP* and *HPRT* gene expression. All quantified gene expression is normalized with No flow – No TGF-β (control) condition.

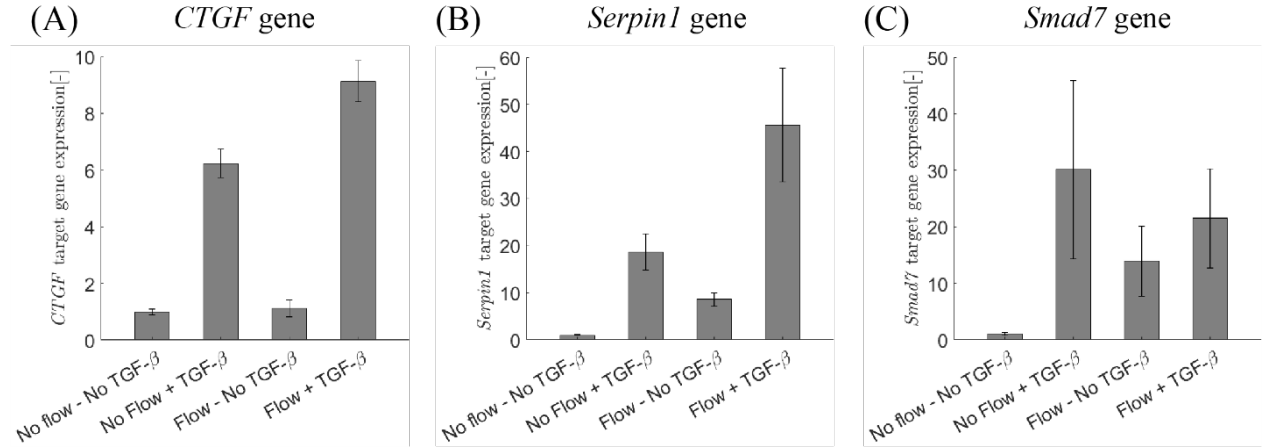

**Figure S6**: **TGF-β target gene analysis on 2D cultured dual-reporter A549 cells in a microfluidic chip subjected to flow (fluid-induced shear stress), exogenous TGF-β and combination of flow and exogenous TGF-β conditions, in contrast with no flow and TGF-β condition (as control condition).** qPCR results are representative of an average of three technical replicates from two independent experiments for No-Flow conditions and four independent experiments for Flow conditions performed in 2D-microfluidics (without matrix).

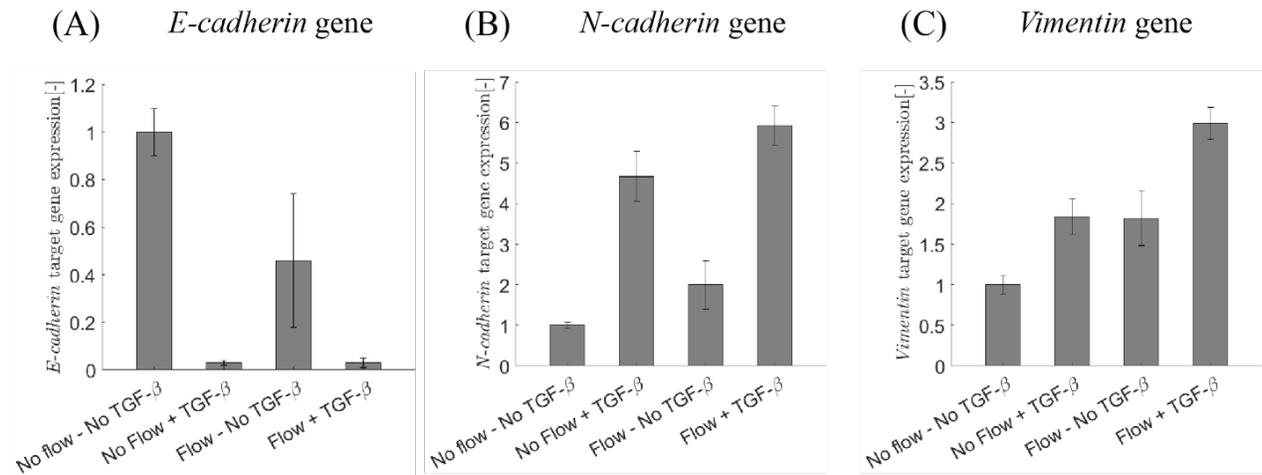

**Figure S7: EMT marker gene analysis on 2D cultured dual-reporter A549 cells in a microfluidic chip subjected to flow (fluid-induced shear stress), exogenous TGF-β and combination of flow and exogenous TGF-β conditions, in contrast with no flow and TGF-β condition (as control condition).** qPCR results are representative of an average of three technical replicates from two independent experiments for No-Flow conditions and four independent experiments for Flow conditions performed in 2D-microfluidics (without matrix).

| Gene | Forward (5' – 3') | Reverse (3' – 5') |
|---|---|---|
| **EMT target genes** | | |
| *E-cadherin* | CCCGGTATCTTCCCCGC | CAGCCGCTTTCAGATTTTCAT |
| *N-cadherin* | CAGACCGACCCAAACAGCAAC | GCAGCAACAGTAAGGACAAACATC |
| *Vimentin* | CCAAACTTTTCCTCCCTGAACC | CGTGATGCTGAGAAGTTTCGTTGA |
| **TGF-β target genes** | | |

| *Smad7* | TCCAGATGCTGTGCCTTCC | GTCCGAATTGAGCTGTCCG |
|---|---|---|
| *CTGF* | TTGCGAAGCTGACCTGGAAGAGAA | AGCTCGGTATGTCTTCATGCTGGT |
| *Serpin1* | CACAAATCAGACGGCAGCACT | CATCGGGCGTGGTGAACTC |
| **Household genes** | | |
| ARP | CACCATTGAAATCCTGAGTGATGT | TGACCAGCCGAAAGGAGAAG |
| HPRT | CTGGCGTCGTGATTAGTGAT | CTCGAGCAAGACGTTCAGTC |

**Table S2: Primer sequence for qPCR gene analysis**

## 7. CAGA-12-GFP and VIM-RFP fluorescent reporter upregulation of A549 spheroids under No flow and IF ($u_m$ = 0.45 µm/s, $\Delta P$ = 30 mbar) conditions at fixed exogenous TGF-β of 10 ng/mL in 3D-matrix based microfluidic chip

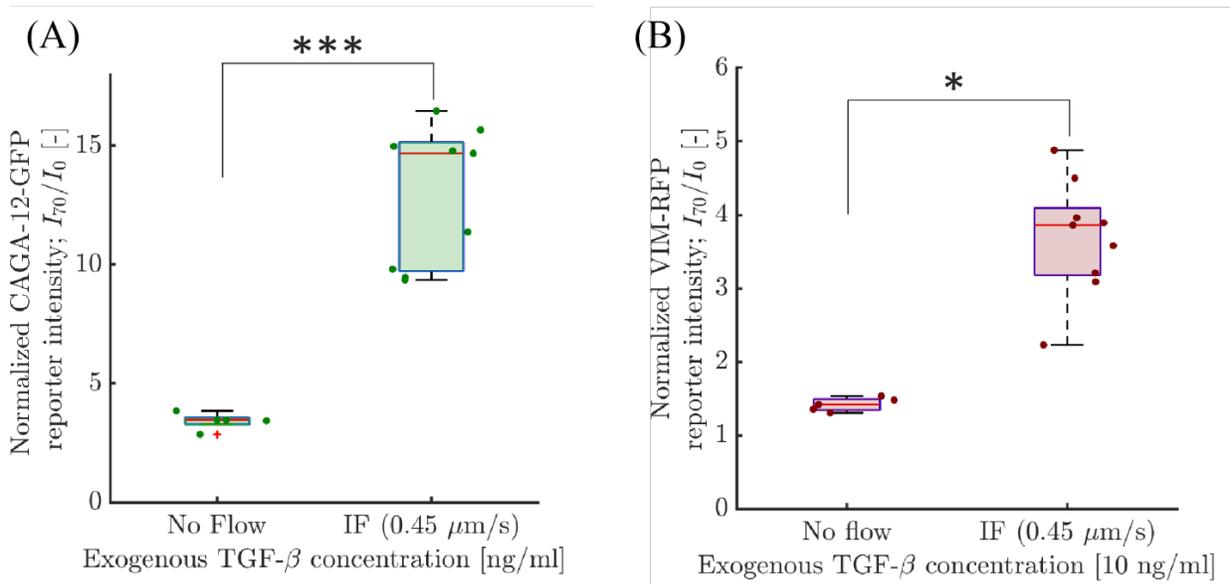

**Figure S8:** A) CAGA-12-GFP reporter and B) VIM-RFP reporter activity when A549 spheroids were exposed to a fixed exogenous TGF-β concentration of 10 ng/mL under no-flow and IF ($u_m$ = 0.45 µm/s, $\Delta P$ = 30 mbar) condition in a 3D-matrix based microfluidic chip.

## 8. Endogenous fluorescent reporter upregulation of A549 spheroids under varying interstitial flow conditions without exogenous TGF-β in 3D-matrix based microfluidic chip

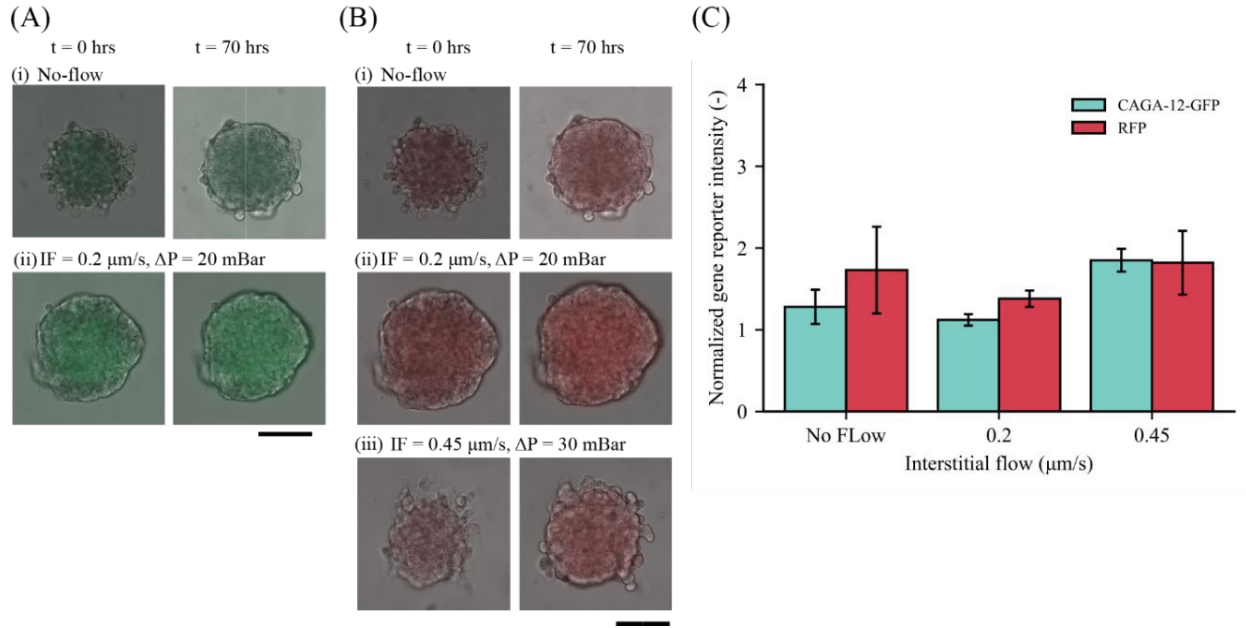

**Figure S9: Endogenous CAGA-12-GFP and VIM-RFP reporter response under varying interstitial flow conditions.** (A) Bright field and GFP channel superposed images of A549 spheroids showing CAGA-12-GFP reporter upregulation at t = 0 and t = 70 hrs for (i) no-flow and (ii) IF ($u_m$ = 0.2 μm/s) conditions. (B) Bright and RFP channel superposed images for vimentin upregulation in no-flow and IF conditions (scale: 100 μm). (C) Normalized CAGA-12G-GFP and VIM-RFP [$I_{70} / I_0$] reporter upregulation at t = 70 hrs.

## 9. CAGA-12-GFP reporter assay using SB-4315422 (a selective small molecule TGF-β type I receptor kinase inhibitor)

### i. CAGA-12-GFP reporter assay in 3D static conditions using A549 spheroids

To test for Smad-dependent CAGA-12-GFP transcriptional reporter activity in the presence of exogenous TGF-β, we performed additional validation studies for the reporter activity using TGF-β type I receptor inhibitor (SB-4315422). Experiments were performed in no-flow (static) conditions using 3D-A549 spheroids embedded in 5wt.% gelMA. A549 spheroids were first treated with SB-431542 inhibitor (10 μM) for 6 hrs before adding exogenous TGF-β. The reporter activity was measured in three different conditions: i) Vehicle control (4 mm HCl in DMSO), ii) SB-431542 (10 μM) treated + exogenous TGF-β (10 ng/mL), and iii) only exogenous TGF-β (10 ng/mL), as shown in Fig. S10. Imaging was performed using Zeiss LSM 980 Confocal Microscope at 10x/NA 0.3 M27 air objective with Z-stack for GFP reporter activity at t = 0 hrs (before TGF-β treatment) and t = 72 hrs (after TGF-β treatment). For GFP fluorescence, laser with the excitation wavelength of 488 nm was used at 1024 x 1024 pixel$^2$ density. For image analysis, Z-stack images were converted to 2D-image using the Z-project function at maximum intensity in ImageJ. The fluorescence intensity at t = 72 hrs of spheroids were normalized with respect to fluorescence intensity at t = 0 hrs. Fig.S10 shows the relative change in expression for fluorescence intensity at t = 72 hrs for all cases (n = 4 spheroids). CAGA-12-GFP transcriptional reporter activity was only observed when A549 spheroids were stimulated only with exogenous TGF-β. On the other hand, A549 spheroids under control does not show any fluorescence upregulation due to inactivity of Smad signaling activity in the absence of exogenous TGF-β molecules. When A549 spheroids were treated with SB-431542 inhibitor, the addition of exogenous TGF-β resulted in no reporter activity. This is due to the inhibition of type I receptor kinase activity preventing Smad-dependent transcriptional reporter response.

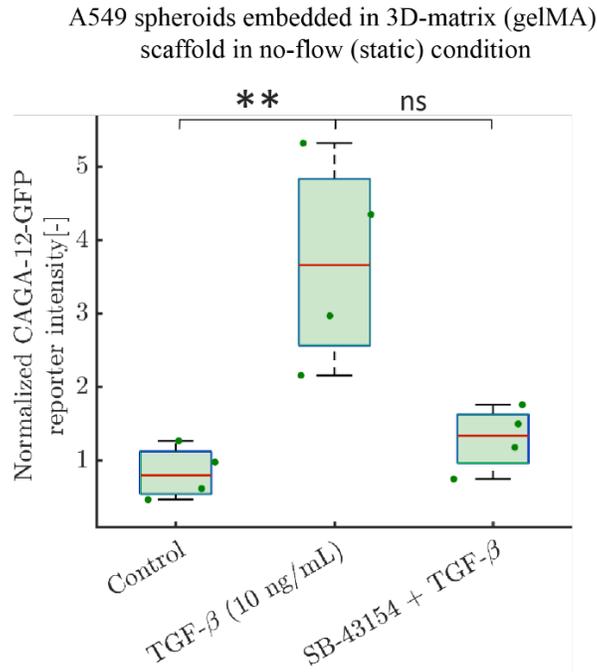

**Figure S10:** CAGA-12-GFP transcriptional reporter activity of A549 spheroids embedded in 5wt.% gelMA (in 3D) at t = 72 hours (n = 4 spheroids for each condition). Reporter expression is highest when stimulated with exogenous TGF-β (10 ng/mL) compared to spheroids first treated SB-431542 (TGF-β receptor kinase inhibitor) followed by exogenous TGF-β (10 ng/mL) stimulation.

ii. **CAGA-12-GFP and VIM-RFP reporter assay in 2D-microfluidics under flow (without exogenous TGF- β) and SB-431542 inhibitor**

To further extend these studies to explore the binding affinity of exogenous TGF- β molecules with TGF-β receptor sites, we performed additional validation studies in 2D-microfluidic device using A549 dual-reporter cells with and without TGF- β inhibitor (SB-431542). 2D-microfluidic devices were prepared by culturing dual-reporter A549 cells 2D in a microfluidic channel according to the method described in section 5 (ESI). In the first experiment, we observed CAGA-12-GFP and VIM-RFP reporter activity of A549 cells cultured in 2D-microfluidics subjected to Flow with and without SB-431542 inhibitor. For inhibitor treatment experiments, after cell-surface attachment, A549 cells were incubated in the presence of TGF-β receptor kinase inhibitor (SB-431542, 10 µM) for 6 hours. Imaging was then performed at t = 0 hr and t = 72 hrs at 4 different locations of the microfluidic channel with imaging conditions described in section 5 (ESI). We plotted the CAGA-12-GFP and VIM-RFP at t = 72 hrs normalized with respect to t = 0 hr. Fig. S10 (A) shows that A549 cells treated with SB-431542 (TGF-β receptor kinase inhibitor) blocks all CAGA-12-GFP reporter activity. Whereas, A549 cells under flow produces significant reporter expression. We observe similar effect in VIM-RFP reporter, where cells treated with SB-431542 inhibitor showed reduced RFP reporter expression compared to cells subjected to flow, shown in Fig S10 (B). This experiment thus highlights the additional effect of flow in 2D towards Smad-dependent transcriptional reporter response.

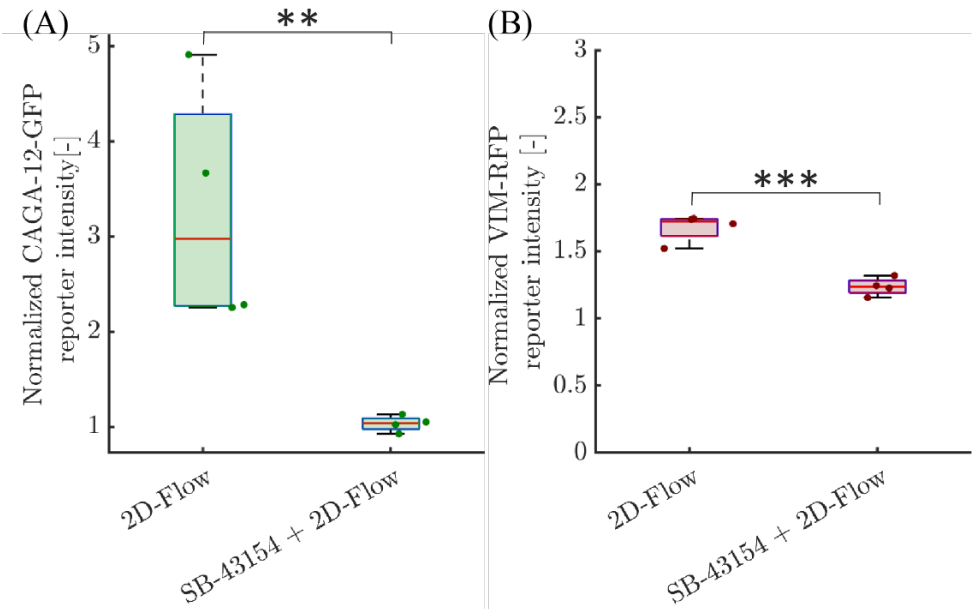

**Figure S11: CAGA-12-GFP and VIM-RFP reporter expression in 2D-microfluidics (without matrix) under flow treated with and without SB-431542 (TGF-β inhibitor).** A) CAGA-12-GFP reporter and B) VIM-RFP expression at t = 72 hrs shows significant reporter upregulation when compared to reporter expression for A549 cells treated with SB-431542 inhibitor.

### iii. CAGA-12-GFP and VIM-RFP reporter assay cultured in 2D-microfluidics under flow conditions with exogenous TGF- β and SB-431542 inhibitor

In the second experiment, A549 cells in 2D-microfluidics, CAGA-12-GFP and VIM-RFP reporter activity was quantified under flow and exogenous TGF-β with and without SB-431542 inhibitor treatment. These experiments further confirm the CAGA-12-GFP reporter activity is a consequence of the exogenous TGF-β molecules binding with TGF-β receptor sites to undergo Smad signaling. In Fig. S11 (A), A549 cells treated with SB-431542 inhibitor blocked all TGF-β receptor kinase activity in the presence of exogenous TGF-β molecules to prevent Smad-signaling response. This led to inactivity of CAGA-12-GFP transcriptional reporter upregulation. On the other hand, significant upregulation of CAGA-12-GFP activity was observed when no inhibitor treatment was performed. Similar response was obtained for VIM-RFP reporter activity, where flow and exogenous TGF-β exposure leads to higher vimentin activity, shown in Fig. S11 (B).

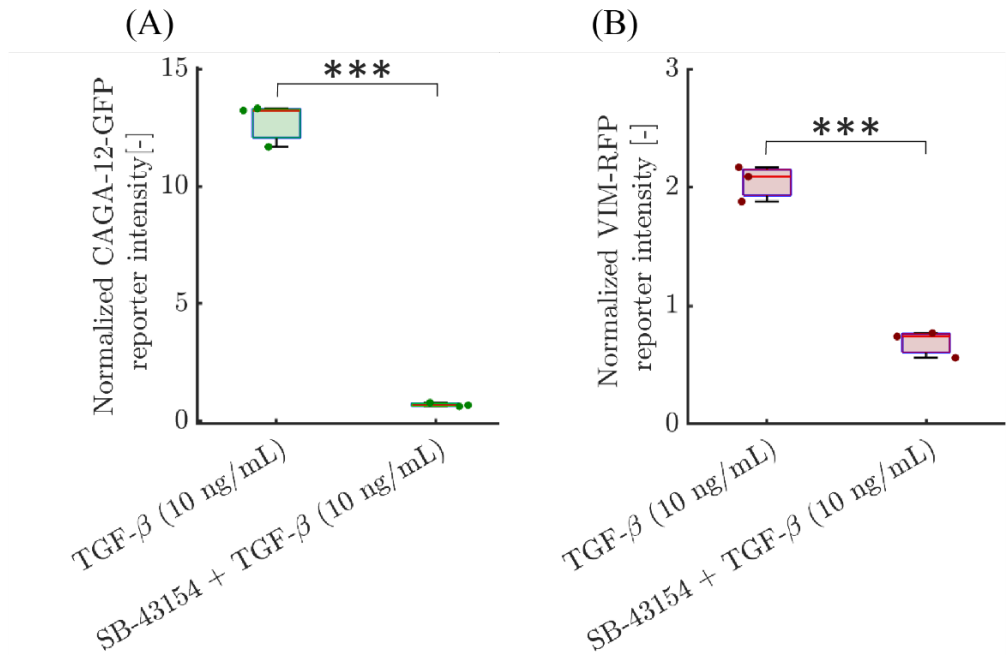

**Figure S12: CAGA-12-GFP and VIM-RFP reporter expression in 2D-microfluidics with flow and exogenous TGF-β conditions treated with and without SB-431542 (TGF-β inhibitor).** A) CAGA-12-GFP reporter and B) VIM-RFP expression at t = 72 hrs shows significant upregulation in reporter expression when compared to reporter expression for A549 cells treated with SB-inhibitor performed in 2D-microfluidics (without matrix).

## 10. CAGA-12-GFP fluorescence profile under no-flow conditions with varying exogenous TGF-β concentrations in 3D-matrix based microfluidics.

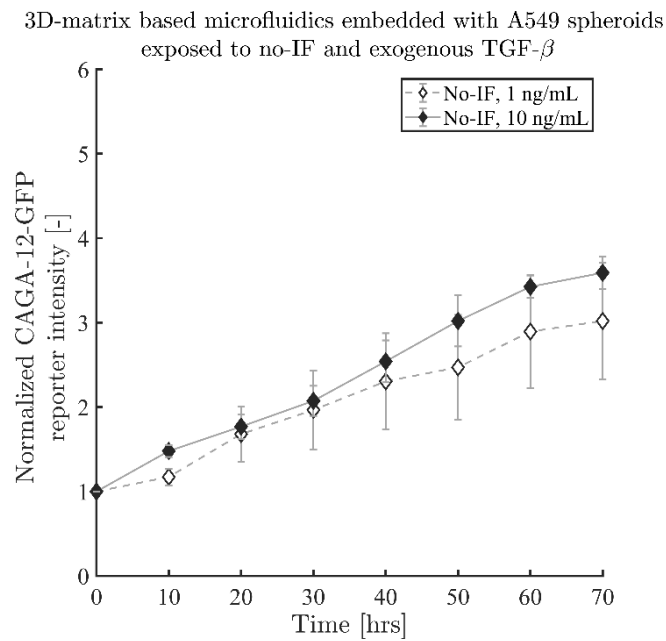

**Figure S13: CAGA-12-GFP fluorescence upregulation profile of A549 spheroids embedded in gelMA under no-IF**

(static) condition supplemented with varying exogenous TGF-β concentration of 1 and 10 ng/mL in 3D-matrix based microfluidic chip.

## 11. Evolution of exogenous TGF-β concentration and surface adsorption under varying interstitial flow condition

Based on the numerically computed velocity and pressure fields (see Section 4), we measured the temporal evolution of TGF-β concentration within the porous hydrogel matrix by solving the mass transport equations (using the Transport of Diluted Species module in COMSOL Multiphysics). To mimic the experimental conditions, the channel AB was initially filled with TGF-β (with molecular diffusivity 21.3 µm$^2$/s) at concentration $C_0$, while the remaining domains were kept at zero initial concentration. For the conditions with interstitial flow, the two inlets A and B were supplied with a continuous injection of TGF-β at concentration $C_0$. For the no-IF condition, the continuous injection condition was removed, and transport was fully dominated by the diffusion of TGF-β from the channel AB to the hydrogel (middle channel) with a spheroid model. The evolution of TGF-β concentration near the spheroid (at a radial distance of 0.05 mm from the interface) for various IF and no-IF conditions are shown in Fig. S10.

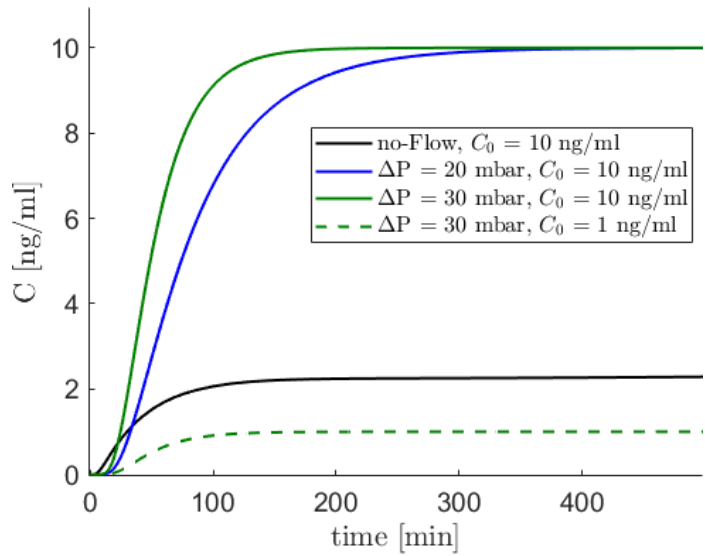

Figure S14: Evolution of average concentration of exogenous TGF-β ($C_0$ = 1 and 10 ng/mL) near a spheroid (within a radial distance of 0.05 mm) at interstitial flow conditions (ΔP = 0 (no-flow), 20 and 30 mbar)

## 12. Visualizing local heterogeneity in fluorescence expression of A549 spheroid stimulated by exogenous TGF-β

To visualize the evolution of fluorescence profile of a spheroid over the 70 hrs time frame, we performed image analysis using the Polar Transformer function (ImageJ plug-in), see Fig.S15 (A). Since a spheroid in general closely represents a circular object (in 2D plane, with xy coordinate), the spheroid can be "unwrapped" by converting this 2D-spheroid image into radius/angle coordinates (polar transformation). We do this for all spheroid images obtained at time points t = 0, 24, 48 and 70 hour, see Fig. S15(B). Using

ImageJ, we measured the averaged fluorescence intensity at every angle (θ = 0 to 360 degrees) of the converted polar transformed image, see Fig. S15(B). Using a custom-made MATLAB script, we plot the average fluorescence intensity (for every θ), see Fig. S15(C).

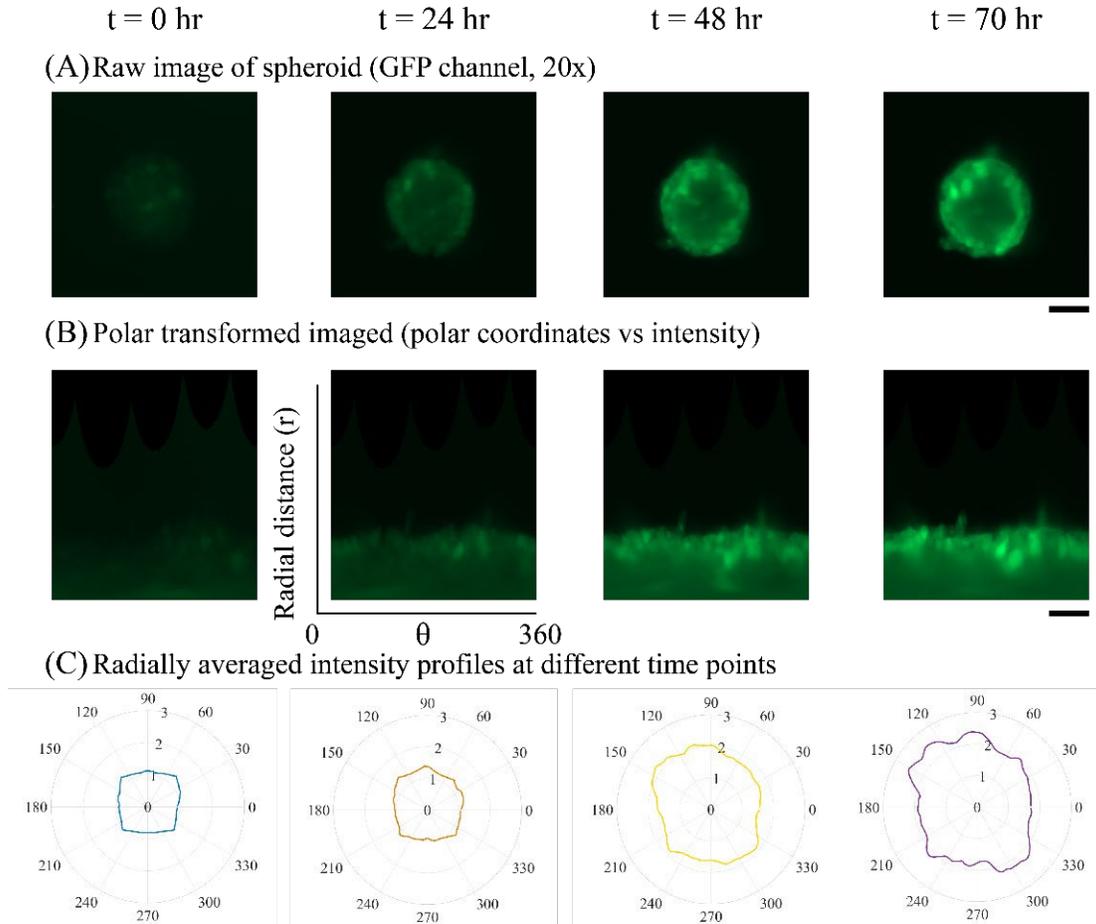

**Figure S15: Methodology for polar plot analysis:** (A) Raw GFP fluorescence channel image (20x) of a spheroid at different time intervals exposed to exogenous TGF-β (10 ng/mL) under no-flow conditions, scale: 100 µm. (B) Cartesian-polar transformed image obtained via Polar transformer plug-in in ImageJ, scale: 100 µm. The spheroid (circular object) is converted into an image represented in polar coordinates (θ=0 to 360 degrees). (C) Polar plot of radially averaged fluorescence intensity at t = 0, 24, 48 and 70 hrs. Each profile is normalized by the intensity at t = 0 hr to show the evolution of signal intensity in radial coordinates.

### Supplementary movies

Supplementary movie S1: Time lapse video of bright-field images of A549 spheroid embedded in gelMA in 3D-microfluidic chip exposed to IF ($u_m$ = 0.45 µm/s, ΔP = 30mbar) and exogenous TGF-β (10 ng/mL). Spheroids show cellular motion activity at spheroid edges over the timeframe of experiment. Most activity is visible in top/side part of the spheroid periphery (in the direction of IF).

Supplementary movie S2: Time lapse video of bright-field images of A549 spheroid embedded in gelMA in 3D-microfluidic chip exposed to no-IF (ΔP = 0mbar) and exogenous TGF-β (10 ng/mL). Spheroids show minimal cellular motion at spheroid edge over the timeframe of experiment.